\documentclass[]{aa} 
\usepackage{natbib}           
\usepackage{graphicx}         
\usepackage{amssymb}          
\usepackage[usenames,dvipsnames,svgnames]{xcolor}
\usepackage{txfonts}

\newcommand{\p}[1]{#1}  

\newcommand{\BE}{\begin{equation}}
\newcommand{\EE}{\end{equation}}
\newcommand{\BA}{\begin{eqnarray}}
\newcommand{\EA}{\end{eqnarray}}
 \newcommand{\fig}[1]{Figure~\ref{fig_#1}}
 \newcommand{\figs}[2]{Figures~\ref{fig_#1} and \ref{fig_#2}}
 
 \newcommand{\sect}[1]{Section~\ref{sect_#1}}
 \newcommand{\sectss}[2]{Sections~\ref{sect_#1}-\ref{sect_#2}}
 \newcommand{\eq}[1]{Equation~(\ref{eq_#1})}
 \newcommand{\eqs}[2]{Equations~(\ref{eq_#1}) and (\ref{eq_#2})}
 \newcommand{\eqss}[2]{Equations~(\ref{eq_#1}-\ref{eq_#2})}

\newcommand{\rmd}{ {\mathrm d} }
\renewcommand{\vec}[1]{ {\bf #1} }
\newcommand{\uvec}[1]{ \hat{\bf #1} }

\newcommand{\curl}{ {\bf \nabla} \times}

\newcommand{\adhoc}{\textit{ad hoc}}

\newcommand{\eg}{\textit{e.g.}}
\newcommand{\etal}{\textit{et al.}}
\newcommand{\ie}{\textit{i.e.}}
\newcommand{\insitu}{in situ}
\newcommand{\commonCap}{The straight lines are linear fits to the data points showing the global tendency.  $\lA >0$ and $\lA <0$ are respectively shown in red and blue, and the abscissa, $|\lA|$, allows the comparison of the two FR sides (\fig{schema}).  The results with the full MC sets are shown in black. $c_p$ and $c_s$ are respectively the Pearson and Spearman rank correlation coefficients. $\mu$ is the mean value of the ordinate and $\sigma$ is the standard deviation of the fit residuals.}

\newcommand{\Aa}{A_{\rm a}}

\newcommand{\Bo}{B_0}  
\newcommand{\Ba}{B_{\rm a}}
\newcommand{\BL}{\vec{B}_{\rm L}}
\newcommand{\Bz}{B_{\rm z}}
\newcommand{\chid}{\chi_{\rm dir}}  
\newcommand{\chiR}{\chi_{\rm R}}  
\newcommand{\cP}{c_{\rm P}}  
\newcommand{\cS}{c_{\rm S}}  
\newcommand{\degree}{^\circ} 
\newcommand{\del}{} 

\newcommand{\dQsdRobs}[1]{{\mathrm d} #1 _{\rm obs.} / {\mathrm d} R}
\newcommand{\dQsdRtot}[1]{{\mathrm d} #1 _{\rm total}/ {\mathrm d} R}
\newcommand{\ea}{\hat{\bf e}_{\rm a}}
\newcommand{\ez}{\hat{\bf e}_{\rm z}}
  
\newcommand{\Ftot}{F_{\rm total}}
\newcommand{\Hmc}{H_{\rm MC,cycle}}
\newcommand{\Htot}{H_{\rm total}}
\newcommand{\iA}{i}
\newcommand{\latmax}{\theta_{\rm max}}
\newcommand{\lA}{\lambda}
\newcommand{\lS}{\lambda_{\rm sup}}

\newcommand{\Lint}{L_{\rm int}}
\newcommand{\Ltot}{L_{\rm total}}
\newcommand{\MCa}{MC$_{\rm Ly}$}
\newcommand{\MCb}{MC$_{\rm Le}$}

\newcommand{\Npart}{N_{\rm part}}
\newcommand{\nt}{n_{\rm t}}
\newcommand{\Ntot}{N_{\rm total}}
\newcommand{\pA}{\phi}
\newcommand{\phimax}{\varphi_{\rm max}}
\newcommand{\pobs}{\mathcal{P}_{\rm obs}}
\newcommand{\pobsl}{\mathcal{P}_{\rm obs}(\lA)}

\newcommand{\pvphi}{\mathcal{P}_{\varphi}}

\newcommand{\Ro}{R}  
\newcommand{\sa}{s}
\newcommand{\tA}{\theta}

\newcommand{\ur}{\hat{\bf u}_{\rho}}

\begin{document}

\title{Magnetic Flux and Helicity of Magnetic Clouds}

\author{P.~D\'emoulin\inst{1} \and
        M.~Janvier\inst{2,3} \and
        S.~Dasso\inst{4,5}      
       }

   \institute{
   Observatoire de Paris, LESIA, UMR 8109 (CNRS), F-92195 Meudon Principal Cedex, France \email{Pascal.Demoulin@obspm.fr}\\
   Department of Mathematics, University of Dundee, Dundee DD1 4HN, Scotland, United Kingdom,\email{mjanvier@maths.dundee.ac.uk} \\
   Institut d'Astrophysique Spatiale, Batiment 121, Univ. Paris-Sud - CNRS, 91405, Orsay Cedex, France\\
   Instituto de Astronom\'\i a y F\'\i sica del Espacio, UBA-CONICET, CC. 67, Suc. 28, 1428 Buenos Aires, Argentina
   \email{sdasso@iafe.uba.ar} \\
   Departamento de Ciencias de la Atm\'osfera y los Oc\'eanos and Departamento de F\'\i sica, Facultad de Ciencias Exactas y Naturales, Universidad de Buenos Aires, 1428 Buenos Aires, Argentina 
\email{dasso@df.uba.ar}  
           }

\date{Received ***; accepted ***}

   \abstract
 {Magnetic clouds (MCs) are formed by flux ropes (FRs) launched from the Sun as part of coronal mass ejections (CMEs).  They carry away an important amount of magnetic flux and helicity.}
  {The main aim of this study is to quantify these quantities from \insitu\ measurements of MCs at 1~AU. }
  {The fit of these data by a local FR model provides the axial magnetic field strength, the  radius, the magnetic flux and the helicity per unit length along the FR axis.}  
  {We show that these quantities are statistically independent of the position along the FR axis. We then derive the generic shape and length of the FR axis from two sets of MCs. These results improve the estimation of magnetic helicity.  
   Next, we evaluate the total magnetic flux and helicity crossing the sphere of radius of 1~AU, centered at the Sun, per year and during a solar cycle.  We also include in the study 
   two sets of small FRs which do not have all the typical characteristics of MCs. }  
{While small FRs are at least ten times more numerous than MCs, the magnetic flux and helicity are dominated by the contribution from the larger MCs.  
They carry in one year the magnetic flux of about 25 large active regions and the magnetic helicity of 200 of them.  MCs carry away an amount of unsigned magnetic helicity comparable to the one estimated for the solar dynamo and the one measured in emerging active regions.
\textbf{Note:} This article will be published in Solar Physics but is set in the A\&A format because of LaTeX compilation problem with SP style.  }


   \maketitle

\section{Introduction}
     \label{sect_Introduction} 

Magnetic clouds (MCs) are a subset of Interplanetary Coronal Mass Ejections (ICMEs) characterised by an enhanced and smooth magnetic field strength, a large and coherent rotation of the magnetic field and a low proton temperature compared with the typical solar wind with the same velocity \citep{Burlaga81}.  They are the continuation in the interplanetary medium of CMEs launched from the solar corona after an instability has occurred in the coronal magnetic field.
Because of its observed properties, the large-scale magnetic configuration of MCs is frequently modeled by a magnetic flux rope (FR).  

Among others, two global quantities characterise a flux rope: its axial magnetic flux $F$ and its magnetic helicity $H$.  
This last quantity quantifies how all the elementary magnetic flux tubes are winded around each other in a defined volume. \p{Magnetic helicity has several remarkable properties both from the theoretical and observational view points \citep[\eg , see the reviews of ][]{Demoulin07, Demoulin09c, Pevtsov14}. In particular,} $H$ is an ideal magnetohydrodynamic invariant that can be obtained from an invariant associated with electrons in a proton-electron multifluid description, 
in the limit of zero electron inertia \citep[see \eg , ][]{Andres14}. In a closed system, magnetic helicity is almost conserved in resistive MHD on 
a timescale lower than the global diffusion timescale \citep{Matthaeus82b,Berger84}, while \eg\ magnetic energy is largely transformed into other forms of energies. \p{This theoretical prediction was tested positively with MHD simulations of coronal jets \citep{Pariat15}.}

The axial magnetic flux $F$ and the magnetic helicity $H$ are conserved during the FR propagation unless the FR significantly 
reconnects with the surrounding solar-wind magnetic field.   
This conservation property was used to quantitatively link FRs observed \insitu\ 
to their solar sources \citep[\eg , ][]{Dasso05,Luoni05,Mandrini05,Qiu07,Rodriguez08,Hu14} and to relate the \insitu\ observations of two spacecraft, 
at 1 and 5.4 AU, of the same MC \citep{Nakwacki11}.
Quantification of $H$ and $F$ also allow us to constrain models of coronal formation and ejection of flux ropes to the interplanetary medium, 
as well as of the dynamical evolution of MCs in the solar wind \citep[for a review see][]{Dasso09b}.

The computations of $F$ and even more of $H$ are challenging because magnetic data are only available along the spacecraft trajectory, so along a 1D cut of the FR, while these global quantities are 2D and 3D, \p{\ie\ they are surface and volume integrals,} respectively.  
Then, their estimation relies on flux rope models with the free parameters of the model typically determined by a least square fit to the \insitu\ data \citep[\eg , ][references therein, and \sect{Hestimation}]{Al-Haddad13}.  
All the models provide an estimation of the magnetic field within a cross section of the FR, so they provide $F$ as well as $H$ per unit length along the axis.   
Then, $H$ can be estimated with an \adhoc\ length of the FR, which is typically in the range $[0.5, 2.5]$~AU for a MC observed at 1~AU (see \sect{Hestimation}). 
This supposes a FR uniformly twisted along its length.

Moreover, some theoretical models have been proposed to describe the global shape of MCs.  They have been compared with in situ observations made by a single spacecraft \citep[\eg , ][]{Marubashi07,Hidalgo12}.  But the reconstruction of the 3D global MC shape from in situ measurements of a single event is not satisfactory because it is an ill-posed problem with no unique solution, and these models contain so many free parameters that generally several solutions compatible with the observations are found. 
In a recent work, \cite{Janvier13} proposed a new method to analyse the global shape of the main axis of MCs, from the statistical distribution of the orientation of a large sample of events. This method was recently used to compare observations with different models for the shape of the MC axis \citep{Janvier15}. They found that an ellipsoidal shape is the model that best fits the data, and got an aspect ratio of $\sim 1.2$ for the ellipse. 


The main aim of this study is to further develop the computation of $H$ for MCs based on a statistical analysis of two sets of MCs.   
In \sect{Observations} we first summarise the type of data used, then the equations needed to derive $F$ and $H$, and finally we summarise our present knowledge of $H$ estimations in MCs.  
In \sect{dependence} we investigate how the main flux rope parameters are function of the curvilinear abscissa along the FR axis.   
In \sect{Length} we propose a new method to estimate the length of MC axis; it is based on a statistical study of two MC sets.  
We next use in \sect{Total_Amount} the results of the previous sections to derive the amount of flux and helicity launched from the Sun per year and over a solar cycle by MCs/CMEs.   
We compare these results with the contribution provided by the much more numerous small FRs detected in the solar wind at smaller scales than MCs, as well as with other solar estimations of magnetic helicities (\eg\ dynamo, emerging active regions, solar wind).  
Finally, in \sect{Conclusion}, we summarise our results and outline future studies needed to improve the global helicity budget.

\section{Observations and Models}
      \label{sect_Observations} 

\subsection{Data Sets} 
      \label{sect_Data-Sets}      

In order to perform a statistical study of the magnetic flux and helicity of MCs, we select the two largest lists of analysed MCs presently available. 
\citet{Lynch05} studied 132 MCs observed nearby Earth by Wind and ACE spacecraft during the period 1995-2003.  
\citet{Lepping10} studied 98 MCs observed by Wind spacecraft. This list was extended to the time period of February 1995 to December 2009 (Table~2 at http://wind.nasa.gov/mfi/mag\_cloud\_S1.html
Removing a few MCs that were badly observed (crossing too close from the boundary), it remains 107 MCs \citep[see ][for more information]{Janvier13}. Below, we refer to MCs of both lists as \MCa\ and \MCb\ sets, respectively.

The local magnetic configuration of the studied MCs was deduced in both studies by following the fitting procedure of \citet{Lepping90}, \ie\ with a least square fit of the magnetic field data along the spacecraft trajectory with a linear force-free magnetic field having a circular section and a straight axis \citep{Lundquist50}. The linear force-free field corresponds to the relaxed state with minimum energy for a given helicity content and axial field distribution. 
In the FR coordinates, with $z$ along the FR axis, the \p{magnetic field $\BL$ of the Lundquist's} model writes
  \begin{equation}  \label{eq_Lundquist}
  \BL = \Bo [J_1(\alpha r) \ea + J_0(\alpha r) \ez ] \,,
  \end{equation}
where $J_0$ and $J_1$ are the ordinary Bessel functions of order $0$ and $1$, and $\ea$ and $\ez$ are the azimuthal and axial unit vectors in cylindrical coordinates. $\Bo$ is the magnetic field strength on the axis and $\alpha$ is the linear force-free constant.
The authors selected the boundaries of the MC such that the magnetic field becomes purely azimuthal there, \ie\ they selected $|\alpha| = c/R$, where the constant $c$ is the first zero of $J_0$ ($c\approx 2.4$) and $\Ro$ is the FR radius.  
The handedness of the FR is defined by a another parameter (which values are $\pm 1$).  It is equivalent to define a signed $\alpha$ parameter.

\begin{figure}  
 \centerline{\includegraphics[width=0.5\textwidth]{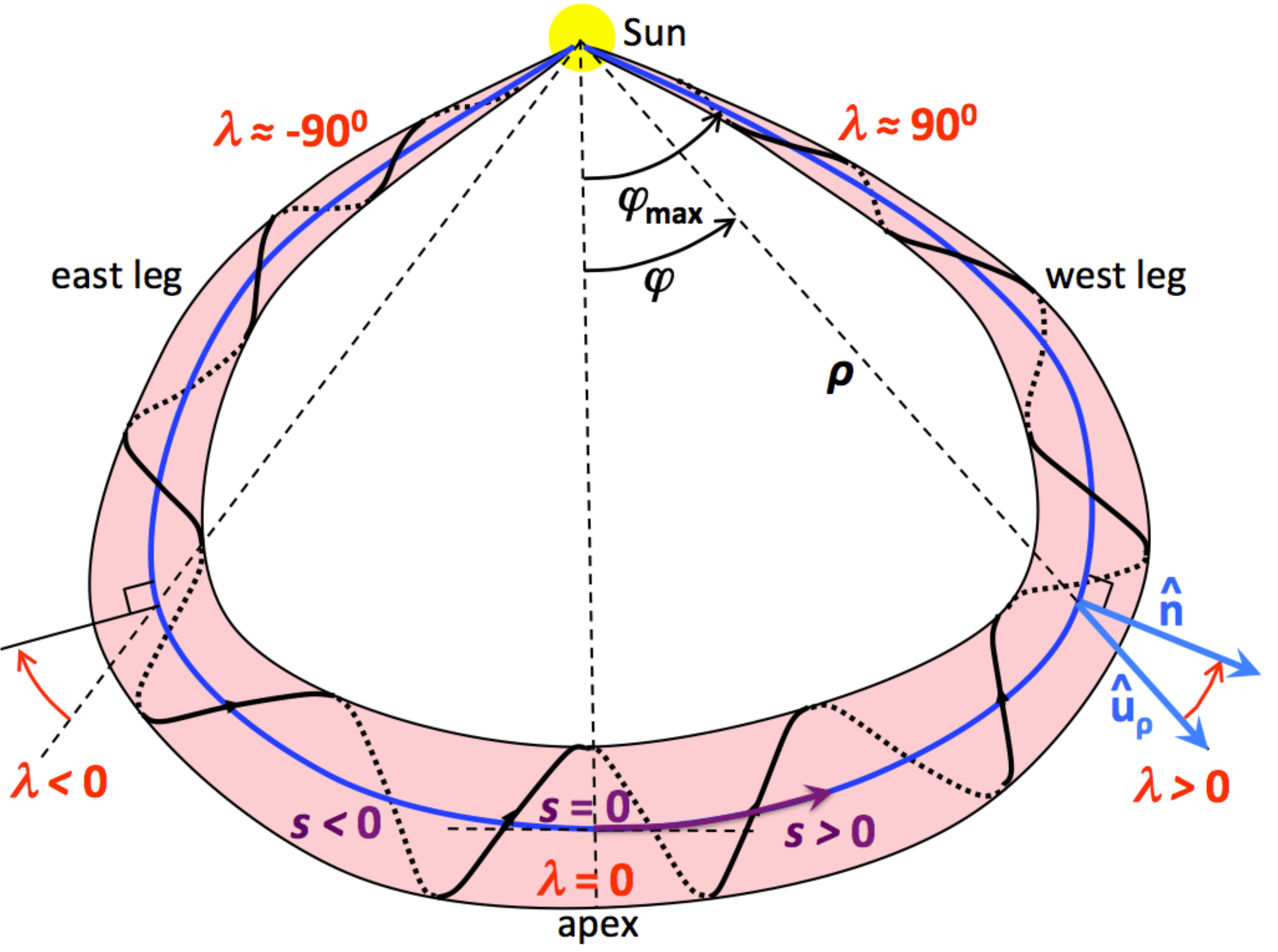} }
\caption{
Schema showing the definition and the large-scale meaning of the location angle $\lA$ for a FR launched from the Sun.  The FR structure is outlined by one twisted field line (black) with dotted style for the behind side.  
The schema shows the plane of the FR axis which is inclined by an angle $\iA$ on the ecliptic.  $\lA$ is defined by the angle between the radial direction [$\ur$] and the normal to the axis [$\uvec{n}$]. The cylindrical coordinates of a point along the axis are ($\rho ,\varphi$).  
The full range of $\varphi$ is 2 $\phimax$.
The signed curvilinear coordinate [$s$] is defined along the FR axis with its origin set at the apex. 
}
 \label{fig_schema}
\end{figure}  

The orientation of the axis is defined by its longitude [$\pA$] and latitude [$\tA$] in the Geocentric Solar Ecliptic (GSE) system of reference.  Another parameter is the closest approach distance [$Y_0$] between the spacecraft trajectory and the MC axis.  It is frequently normalised to the FR radius, and called the impact parameter [$p=Y_0/\Ro$].   
Taking into account the observed mean velocity inferred from the \insitu\ plasma measurements, the least square fit of the above model to the magnetic data determines the six parameters: $\Bo$, $\Ro$, $\pA$, $\tA$, $p$ and ${\rm sign} (\alpha)$.  The fit is done in two steps.  In the first step, each magnetic field vector is divided by its norm (to avoid a bias due to a typical asymmetry of field strength between the front and rear of MCs). In the second step, a fit to the data is realised to determine $\Bo$ (keeping the other parameters fixed).
The quality of the MC fit is measured by the square root of the chi-squared $\chid=\sqrt{\chi^2/N_{\rm d}}$ for \citet{Lynch05} and 
by the square root of the reduced chi-squared defined as $\chiR=\sqrt{\chi^2/(3N_{\rm d}-n)}$ for \citet{Lepping10} where $n=5$ is the number of parameters of the fit and $N_{\rm d}$ is the number of data. 
Both $\chid$ and $\chiR$ are computed during the first step of the fit: they are dimensionless quantities, and measure how well the model fits the direction of the observed magnetic field.

We also analyse two other lists of interplanetary FRs. They can be found in the papers of \citet{Feng07} and \citet{Feng08}.  The detected FRs were also fitted with the Lundquist field with a similar procedure as \citet{Lepping90}, and the derived list of events contains mostly the same parameters (see the above papers for the small differences). 
The list of \citet{Feng07} has $144$ FRs with both MCs and small FRs, while the list of \citet{Feng08} contains $125$ small FRs. 
  
\citet{Janvier14} have analysed the distributions in MC radius of the four data sets and found that small FRs and MCs have different distributions: a power-law for small FRs and a Gaussian-like distribution for MCs. They concluded that small FRs have different solar origins compared with MCs.  We refer to this study for further information on the data sets.  

\citet{Janvier13} have introduced a new spherical coordinate system for the FR axis since the longitude $\pA$ has a large error for FR oriented close to the north-south direction.  They set the polar axis of the new spherical coordinate system along $\uvec{x}_{GSE}$, then they defined the inclination on the ecliptic [$\iA$] and the location [$\lA$] angles.  $\lA$ is defined by the angle between the radial from the Sun and the normal to the axis (\fig{schema}).  For a FR with a known axis shape, of the type shown in \fig{schema}, the angle $\lA$ has a monotonous variation along the axis.  For these cases, $\lA$ labels the location along the axis where the spacecraft intercepts the FR.  The sign convention of $\lA$ is such that, for a FR close to the ecliptic ($\iA$ small) the eastern (western) leg,  corresponds to $\lA<0$ ($\lA>0$). This sign convention is extended to all $\iA$ values by continuity.

\subsection{Theoretical Estimations of Global Quantities} 
      \label{sect_Global}      

Two main global quantities of a FR are its axial magnetic flux [$F$] and its magnetic helicity [$H$].  These global magnetic quantities can only be estimated from the fit of \insitu\ data by a FR model. $H$ is first estimated per unit of length along the FR axis, and it is typically given for a fixed length, which values are discussed in \sect{Hestimation}.

Below we write the expressions of $F$ and $\del H$ for a magnetic field having a cylindrical symmetry, a local approximation for the FR of MCs. Then, $\vec{B}(r)= \Ba(r) ~\ea + \Bz(r) ~\ez$, where $\Ba , \Bz$ are the azimuthal and axial components depending only on the radial coordinate $r$.
Next, we write the specific results for the Lundquist model (\eq{Lundquist}).    

The axial flux, integrated from the axis to the FR radius $\Ro$ and assuming a cylindrical symmetry, is 
  \begin{equation}  \label{eq_FzL}
  F = \int_{0}^{R} \Bz (r') \, 2 \pi \,r' \,\rmd r'
    = \frac{2 \pi J_1(c)}{c} \; \Bo R^2  \,,
  \end{equation}
where the constant $c$ is the first zero of the Bessel function $J_0(r)$.
  
The relative self magnetic helicity of a flux rope is the sum of its twist and writhe helicities \citep{berger06}.  For MCs it is mostly limited to the twisted helicity since the FR axis is thought to have a low writhe as shown for a few MCs observed by several spacecraft \citep[\eg , ][]{Burlaga90,Ruffenach12}. In terms of order of magnitude, the writhe contribution is of the order of 0.1 equivalent turn or below, while the twist is important, of the order of 10 turns. Then, one can consider that the MC helicity is mostly due to the twist.  The helicity [$\del H$] of a straight flux rope of length $L$ is \citep{Berger03b,Dasso03}:
  \begin{equation}  \label{eq_HL}
  \del H = \del L \int_{0}^{R} 2 \Aa (r) \,\Ba (r) \, 2 \pi \, r \, \rmd r 
         = \frac{2 \pi  J_1^2(c)}{c} \; \Bo^2 R^3 \del L \,.
  \end{equation}

  For a FR configuration, magnetic helicity is directly related to the mean number of turns per unit length [$\nt$] of the magnetic field lines along the axis (\sect{Mean-Twist}).  More precisely:
  \begin{equation}  \label{eq_Hnt}
    \del H  = \nt ~F^2 ~\del L\,, 
  \end{equation}
where $\nt$ is a flux weighted mean of the number of turns per unit length along the axis,
  \begin{equation}  \label{eq_ntL}
  \nt = \frac{2}{F^2} \int_{0}^{F} n(F') F' \rmd F' 
      = \frac{c}{2\pi R}   \,, 
  \end{equation}
where $n(F')$ is the local number of turn, which can be expressed in term of the cumulated flux $F'(r)$ from the origin of the flux rope up to a radius $r$, as shown in \sect{Mean-Twist}, and where the right-hand side expression is for the Lundquist model.

\subsection{Helicity Estimation of Magnetic Clouds} 
      \label{sect_Hestimation}

Since magnetic helicity is intrinsically a 3D quantity \p{(\ie , computed with a volume integral)} while observations are limited to the magnetic field measured along the spacecraft trajectory, the estimation of MC helicity involves hypotheses and models.

Within the cylindrical hypothesis, different MC models have been proposed. 
For example, a uniformly twisted field \citep{Dasso03}, a non-force-free field with constant current \citep{Hidalgo00} or with an azimuthal component of the current depending linearly on the radius \citep{Cid02}.  The fit of these models to data introduces a variation of the deduced helicity up to 30\%, which still remains small compared to the variation of helicity computed between different MCs \citep{Gulisano05}. 
Extensions to elliptical cross-section \citep[\eg , ][]{Vandas03} increase the helicity approximately proportionally with the aspect ratio of the cross section \citep{Demoulin09b}.  With typical values of 2 to 3 of the aspect ratio \citep{Demoulin13}, this increases significantly the estimation of the helicity values.  Finally, non force-free models \citep[\eg , ][]{Mulligan01,Hidalgo11,Isavnin11} have also been developed.  
It would be worth to develop both their helicity estimations and their applications to a larger number of MCs using these models, but it is out of the scope of the current paper. 

  The above models have a straight axis configuration. Then, the derived helicity is only a local estimation per unit length along the axis.  
An appealing approach is to extend the above models to toroidal geometry in order to include the curvature of the FR axis \citep[\eg , ][]{Marubashi07,Romashets09}.  Due to a larger number of free parameters, it is not yet clear if they can all be constrained by the data of a single spacecraft.  Two well separated spacecraft provide more constraints to the toroidal model \citep{Nakagawa10}. However, the number of MCs observed is very limited in this configuration as it requires a FR oriented close to the ecliptic plane, where spacecraft are typically located \citep[see the review of ][]{Kilpua11}.
These models also assume an invariance along the curved axis.  In fact, it is not known how the twist is distributed along the MC axis.   We estimate this dependence from a statistical study in \sect{dependence}.

Most MCs are faster than the local solar wind, at least close to the Sun.  It results in the formation of a sheath before the MC where plasma and magnetic field accumulate.
When magnetic fields of different orientation are pushed together, it generally implies magnetic reconnection. This phenomenon can also occur at the rear of the MC, for example, when a faster MC or a fast solar wind stream takes over the propagating MC. This leads to a FR progressively peeling off and only the central region remains as a coherent FR when observed \citep{Dasso06}.  This is confirmed by the presence of magnetic discontinuities \citep{Dasso07,Nakwacki11} and by \insitu\ reconnection signatures \citep{Ruffenach12,Ruffenach15}.  The amount of reconnected flux is case-dependent and significant: it was found that about 40 \% of the total azimuthal magnetic flux on average is lost from this erosion process.  
This process of erosion is typically not taken into account in the definition of MC boundaries, 
in particular for the lists of MCs of \citet{Lynch05} and \citet{Lepping10}.  As such, the helicities estimated in the present paper are expected to be intermediate between those of the FR remaining at 1 AU and of the FR before erosion. 

The reconnection of the FR with open solar wind field can also occur in one leg without direct consequences for the \insitu\ magnetic measurements (when the magnetic perturbation has no time to travel to the crossing location).  However, the tails of the electron distributions provide clues on the large scale connectivities since the faster electrons are fewly interacting with the plasma.  The presence of bi-directional, or counter-streaming,  electron heat fluxes in a MC is generally interpreted as a connection to the Sun at both field-line ends \citep{Richardson91,Shodhan00}.  The counter-streaming electrons are typically present in fragmented and partial (from 0 to 100\%) portions of MCs observed at 1~AU \citep{Shodhan00}.  The counter-streaming electrons in MCs observed at about 5~AU are also very case-dependent, and they are present on average 55\% of the time, a result comparable to observations at 1~AU, so that the amount of disconnection from the Sun does not increase with distance \citep{Crooker04}.  

The length of field lines can be inferred in exceptional cases, when high-energy electrons, accelerated close to the Sun, are injected in them and detected \insitu .  A first method is to derive the path length from the velocities and the different arrival times of electrons of various energies \citep{Kahler06,Masson12}. The second method is based directly on the travel time, so it requires an estimation of the solar release time (with the onset of type III radio bursts) and the \insitu\ detection of the same electron beam.  The unique MC analysed by \citet{Larson97} was recently extended up to a list of 18 MCs \citep{Kahler11,Hu14,Hu15}.  The last studies show an estimation of the length across the flux ropes consistent with what is expected for a flux rope with a more uniformly distributed twist across the cross-section than what is predicted by the Lundquist's model.   All these studies give a large range of lengths, from below 1 to 4~AU.  Since only a few field lines can be probed in one FR leg at most, estimating the flux rope axis length is therefore rather limited.  These results would need to be interpreted with an estimation of the location of the spacecraft crossing (so of the values of the location angle $\lA$ and impact parameter $p$), and with the knowledge of which FR leg the energetic electrons were traveling in (i.e., the sign of $\lA$).   Then, the estimation of the whole effective FR length is not straightforward from these results.  
 
In practice, for the magnetic helicity estimation of MCs, a length is typically assumed. Different values have been used, ranging from 0.5 to 2.5~AU \citep{DeVore00,Lynch05}.  For some specific cases, physical arguments have been invoked to justify the selected length such as the initiation of the solar ejection by the kink instability \citep{Nindos03}, or the disappearance of the solar source region \citep{Mandrini05}, or the agreement between the azimuthal flux estimated in the MC and the flux swept by the flare ribbons in the source region \citep{Du07,Hu14}.

In conclusion, the axis length of the FR is still a major source of uncertainty to estimate MC helicity.  Furthermore, it is not known how helicity is distributed along the MC axis.  The aim of the next sections is to improve our knowledge on these aspects.

\section{Dependence Along the Flux Rope Axis} 
      \label{sect_dependence}

In this section we test whether the MC properties are variable along the MC axis, so whether there is a statistical dependence with the location angle $\lA$ since this angle is also a coordinate along the axis with $|\lA|$ increasing away from the apex (\fig{schema}).
We characterize the correlations by two coefficients: the Pearson [$\cP$] and the Spearman rank [$\cS$] correlation coefficients.  
We report in different figures the fit of the data by a linear function to show global tendencies, as well as the mean value of the studied property [$\mu$] 
and the standard deviation of the fit residuals [$\sigma$] (computed with respect to the fitted straight line). 

\begin{figure*}  
 \centerline{\includegraphics[width=0.7\textwidth]{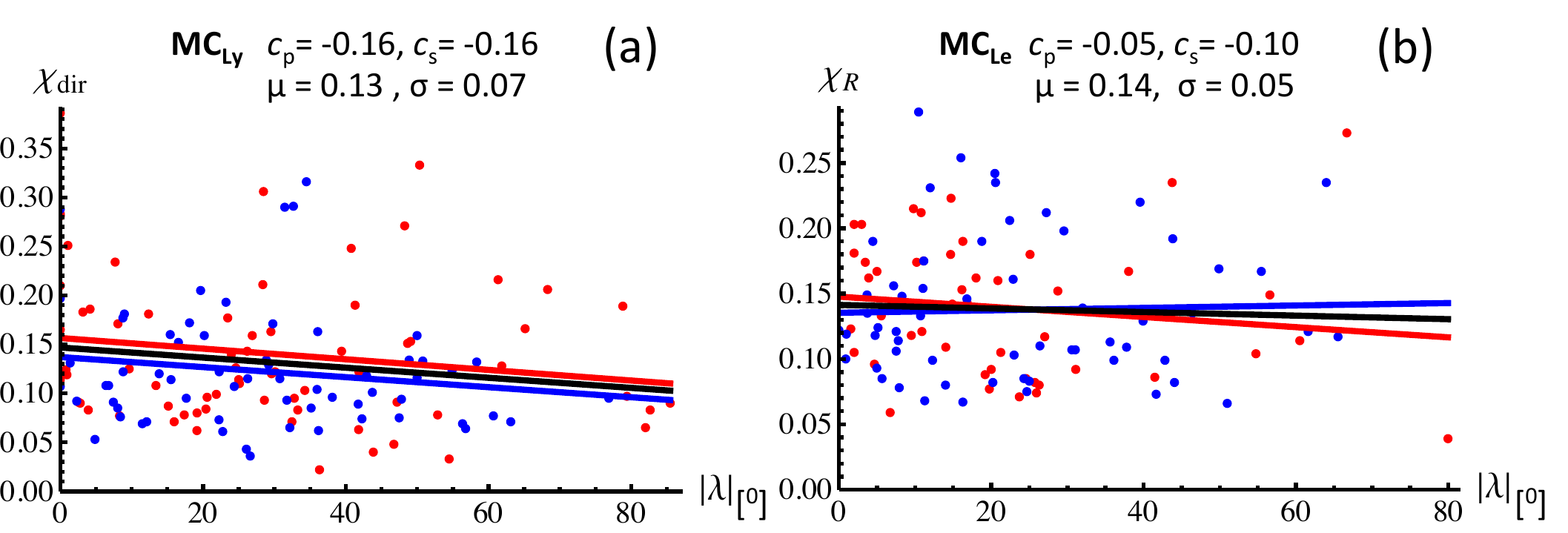} }
\caption{Dependence of the fit quality measured by $\chid$ or $\chiR$ (non dimensionalised) versus the absolute value of the location angle [$\lA$]. 
(a) and (b) are showing MCs analysed by Lynch et al. (2005) and Lepping and Wu (2010), respectively. 
  \commonCap  
  }
 \label{fig_correl_chi}
\end{figure*}  

\begin{figure*}  
 \centerline{\includegraphics[width=0.7\textwidth]{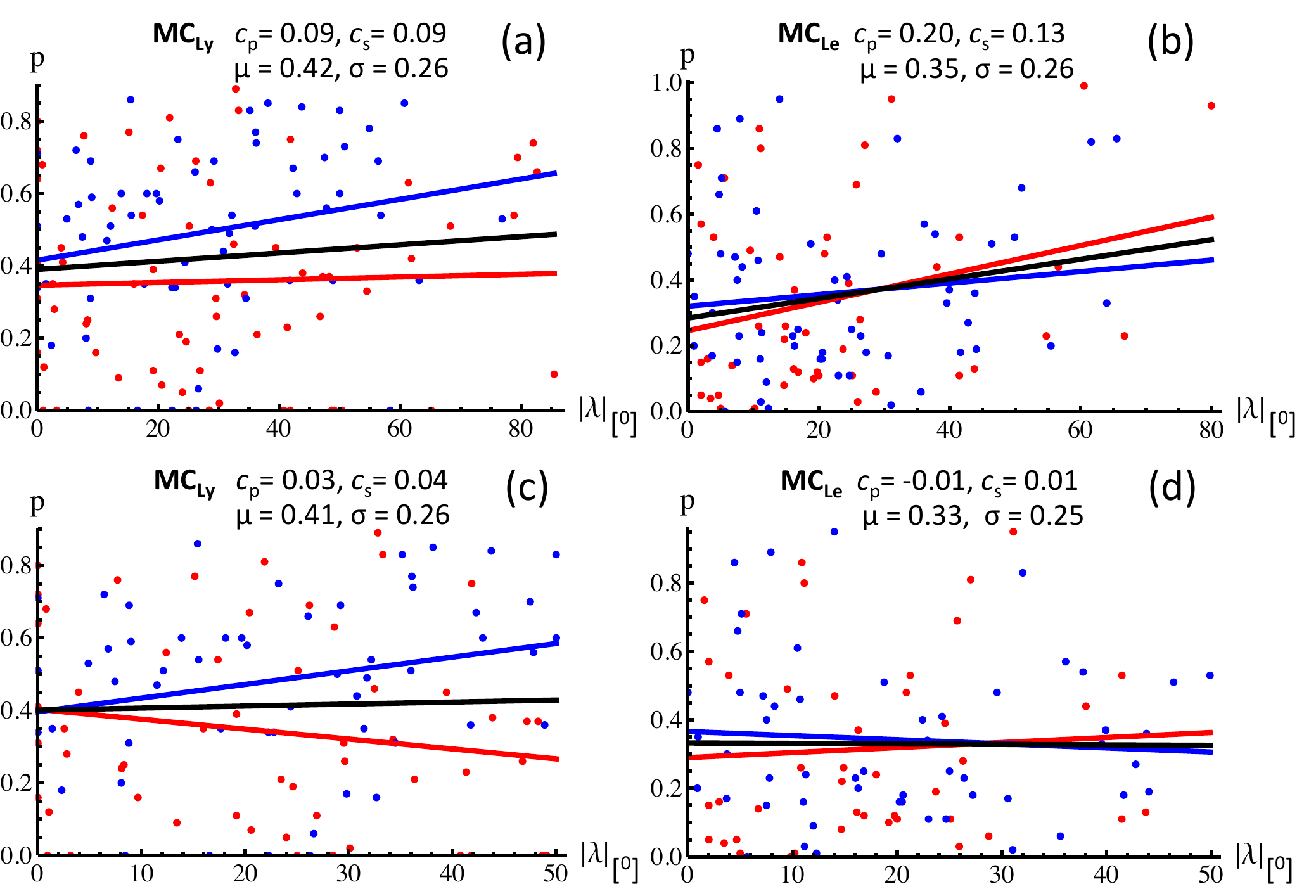} }
\caption{ Dependence of the impact parameter [$p$] versus the absolute value of the location angle [$\lA$] for the two sets of MCs: (a,c) \MCa\ and (b,d) \MCb\ sets.
The top row is for all MCs while the bottom row is for a reduced interval of $|\lA|$ ($< 50 \degree$).
\commonCap
}
 \label{fig_correl_p}
\end{figure*}  

\subsection{Influence of the Spacecraft Trajectory Location} 
      \label{sect_Trajectory}
      
The parameters $\chid$ and $\chiR$ are both testing how well the magnetic field direction of the model fits the \insitu\ data (\sect{Data-Sets}).
For the \MCa\ set, there is a weak tendency ($\cP=\cS=-0.16$) of a lower $\chid$ as $|\lA|$ increases for both MC legs (\fig{correl_chi}a).   
This tendency becomes even weaker when the MCs, crossed near their outer boundaries, are removed from the sample. 
For example, $\cP=\cS=-0.07$ with the selection $|p|<0.7$ where $p$ is the impact parameter. For the \MCb\ set there is no significant correlation $\chiR(\lA)$ for both MC legs (\fig{correl_chi}b).  This result is robust as it is also valid for sub-groups of MCs (\eg\ $\cP=0.04,~\cS=0.06$ with the selection $|p|<0.7$) and there is no significant differences between both legs.   We conclude that the quality of the Lundquist fit to the data is independent of the spacecraft crossing location along the flux rope.
   
The impact parameter $p$ is spread in the interval $[0,1]$ as expected with random distance encounters. Still, low $p$ values are significantly more numerous (\fig{correl_p}).  This is a consequence of the oblateness of the flux-rope cross section \citep{Demoulin13}.
The correlations of the impact parameter $p$ with $\lA$ are positive for \MCa\ and \MCb\ sets (\fig{correl_p}a,b) and all correlation coefficients are small with the selection $|\lA|<50 \degree$ ($|\cP|,|\cS| \leq 0.04$ for \MCa\ set and $|\cP|,|\cS| = 0.01$ for \MCb\ set, \fig{correl_p}c,d).
We interpret this change of $p$ in the flux rope legs with an observational bias, as follows.
As the spacecraft trajectory is crossing the flux rope less perpendicular to its axis (larger $|\lA|$), the spacecraft trajectory explores a longer part along the flux rope.  There, the bending of the MC axis affects the measurements of the magnetic field. It implies that the hypothesis of a local straight flux rope, used in the Lundquist model, is less valid as $|\lA|$ increases \citep{Owens12}. Indeed, for moderate $|\lA|$ values, the deviation between a curved and straight flux rope is small but it becomes strong for large $|\lA|$ values (see their Figures 3-5).
This deviation is interpreted by the Lundquist's fit as a larger impact parameter so that $p$ is positively correlated with $|\lA|$ in \fig{correl_p}a,b.

\begin{figure*}  
 \centerline{\includegraphics[width=0.7\textwidth]{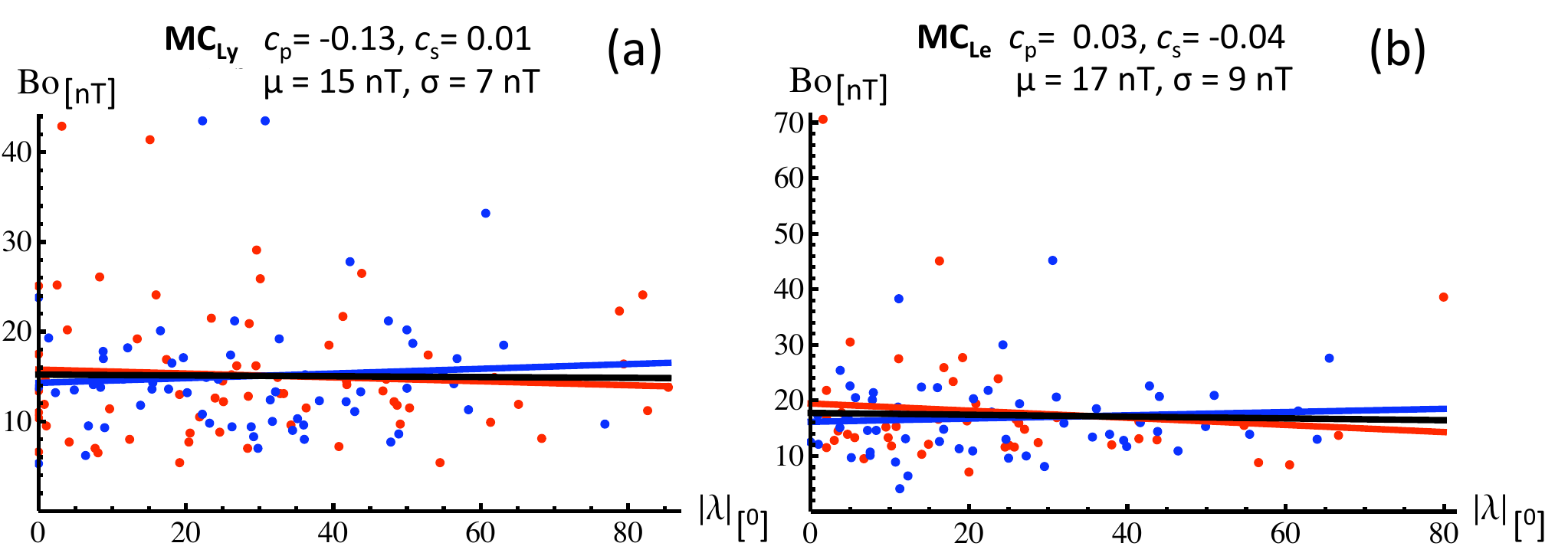} }
\caption{ Dependence of the axial field strength [$\Bo$] 
versus $|\lA|$.
(a) and (b) are showing MCs analysed by Lynch et al. (2005) and Lepping and Wu (2010), respectively.  
  \commonCap 
  }
 \label{fig_correl_Bo}
\end{figure*}  

\begin{figure*}  
 \centerline{\includegraphics[width=0.7\textwidth]{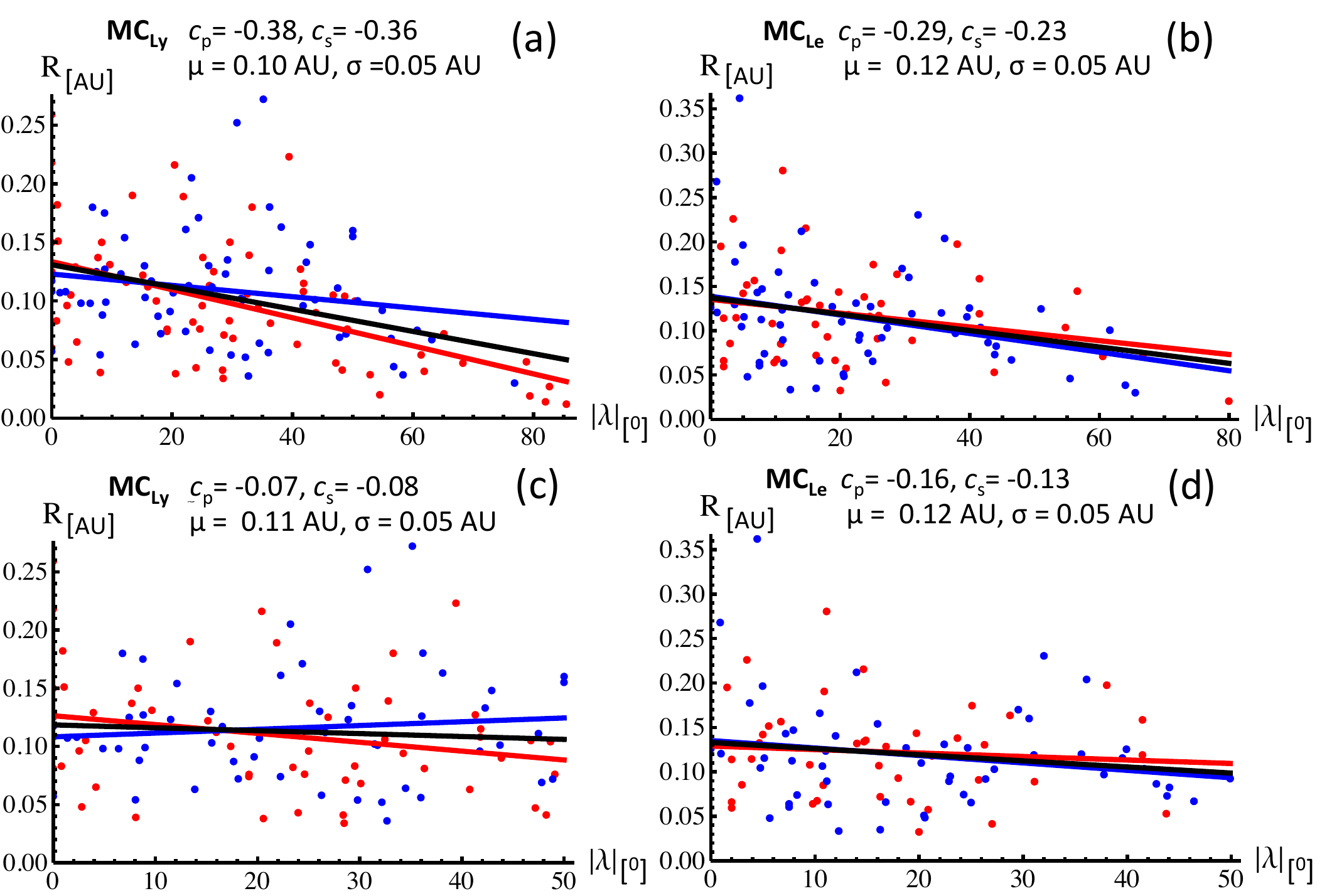} }
\caption{Dependence of the flux-rope radius [$\Ro$] 
versus the absolute value of the location angle [$\lA$] 
for the two sets of MCs: (a,c) \MCa\ and (b,d) \MCb\ sets.
The top row is for all MCs while the bottom row is for a reduced interval of $|\lA|$ ($< 50 \degree$).
  \commonCap }
  \label{fig_correl_R}
\end{figure*}  

\begin{figure*}  
 \centerline{\includegraphics[width=0.7\textwidth]{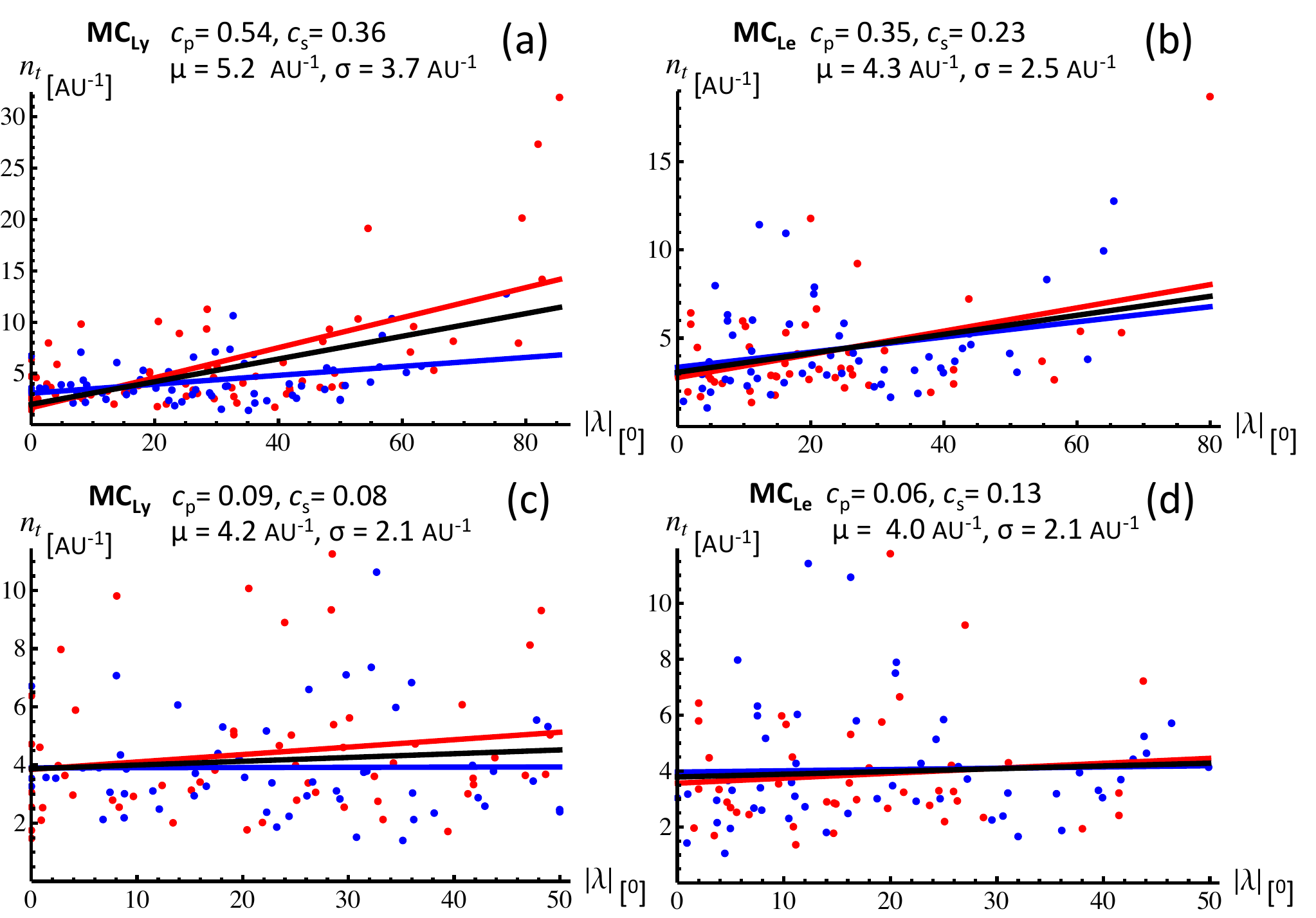} }
\caption{Dependence of the number of turns per unit length [$\nt$] 
versus the absolute value of the location angle [$\lA$] 
for the two sets of MCs: (a,c) \MCa\ and (b,d) \MCb\ sets.
The top row is for all MCs while the bottom row is for a reduced interval of $|\lA|$ ($< 50 \degree$).
  \commonCap }
 \label{fig_correl_nt}
\end{figure*}  

\subsection{Variation of the Physical Parameters Along the Axis} 
      \label{sect_Physical}
      
For both \MCa\ and \MCb\ sets there is no significant correlations between the axial field strength $\Bo$ with $\lA$ (\fig{correl_Bo}a,b).   This result is robust as it stays for sub-groups of MCs.  For example, $\cP=\cS=-0.1$ for \MCa\ set and $\cP=-0.08,~\cS=-0.06$ for \MCb\ set with the selection $|p|<0.7$.  
It also implies that when observed at 1~AU the axial field strength has no significant dependence along the MC axis.  
This result seems a priori contradictory to the standard picture of a MC (\eg\ Figure 1 in \citealt{Richardson10}), for which $\Bo$ would be stronger at the leg than at the apex of the MC. 
The result can then be understood as follows.  MCs strongly expand as they move away from the Sun as a consequence of the approximative balance of total (magnetic and plasma) pressure between the MC and the surrounding solar wind \citep{Demoulin09}.  
This implies that $\Bo$ is mainly a function of the solar distance. 
Then, we interpret the above uniform distribution of $\Bo$ along the MC axis as a consequence of an approximative pressure balance at a fixed observation distance (at 1~AU). Finally, the \MCa\ and \MCb\ sets have comparable $\Bo$ distributions, with a mean value and dispersion of $16 \pm 9$~nT.

In contrast to $\Bo$, $\Ro$ is statistically a decreasing function of $|\lA|$ (\fig{correl_R}a,b) with a stronger anti-correlation for the \MCa\ than for the \MCb\ set.  This anti-correlation is due to the absence of large $\Ro$ values for large $|\lA|$ as can be seen by the absence of blue/red dots in \fig{correl_R}a,b.   Indeed, with the selection $|\lA|<50\degree$, the correlations are much weaker for both sets of MCs (\fig{correl_R}c,d). The correlations are even weaker if the stronger criteria $|\lA|<40\degree$ is applied ($\cP=-0.001,~\cS=-0.04$ for \MCa\ set and $\cP=-0.07,~\cS=-0.06$ for \MCb\ set).  A selection on $|p|$ has a lower effect on the correlations. We interpret these lower values of $\Ro$ in the flux rope legs with an observational bias, as above for the impact parameter $p$. Next, the difference between both legs in \fig{correl_R}c is small and not confirmed by \fig{correl_R}d. Finally, the \MCa\ and \MCb\ sets have comparable $\Ro$ distributions with a mean value and dispersion of $0.12 \pm 0.05$~AU.

We have performed a comparable analysis with the mean number of turns per unit length [$\nt$].   With a Lundquist model, with an axial field vanishing at the flux-rope boundary, $\nt$ is directly related to $\Ro$ (\eq{ntL}). Still, as $\nt (\lA)$ tells how the twist is distributed along the flux ropes axis, we also show the results with $\nt$.  With both the \MCa\ and \MCb\ sets, $\nt$ has a strong positive correlation with $|\lA|$ (\fig{correl_nt}a,b), which is surprising as the MC legs would be more twisted than the apex.  In fact, with the selection $|\lA|<50\degree$, the correlations are weak for both sets of MCs  (\fig{correl_nt}c,d). They are even weaker with the more stringent condition $|\lA|<40\degree$ ($\cP=0.09, ~\cS=0.04$ for \MCa\ set and $\cP=0.01,~\cS=0.06$ for \MCb\ set).   We conclude that the flux ropes are 
uniformly twisted along their axis, at least in the range $|\lA|<50\degree$ around the apex within the limits of the variations between MCs: $\nt = 4 \pm 2$~AU$^{-1}$.
The mean number of turns found here is in agreement with previous studies. For example, \citet{Farrugia09} studied a small and hot flux rope assuming a constant twist model (i.e., the Gold and Hoyle model) and found a number of turns of $\sim 7$ ~AU$^{-1}$. On the other hand, \cite{Mostl09} studied one MC from \insitu\ observations made with two spacecraft (STEREO and Wind) crossing different parts of the cloud, and modeling the magnetic topology with a Grad-Shafranov equilibrium. They found a small variation of the number of turns across the flux rope, with a mean value $\sim 2$ ~AU$^{-1}$.

The above absence of significant correlation for $\Bo$, $\Ro$ and $\nt$ with $\lA$, at least for  $|\lA|<50\degree$, implies also that the global quantities $F$ and $\del H$, \eqss{FzL}{HL}, are also almost independent of $\lA$.   The correlations of $F$ and $\del H$ with $\lA$ can also be done directly.  However, the main limitation of this approach is the much broader range of variation within a MC set since non linear dependencies on $\Bo$ and $\Ro$ are present in $F$ and $\del H$ (\eqss{FzL}{HL}) and there is also a positive correlation between $\Bo$ and $\Ro$ ($\cP=0.39, ~\cS=0.35$ for \MCa\ set and $\cP=0.31,~\cS=0.23$ for \MCb\ set). This implies a much larger dispersion of these global quantities so that a correlation study is less pertinent (\eg\ it is more affected by outsiders).  This is especially true for $\del H$, which has the strongest non-linearities, while, in contrast, the mean number of turns [$\nt$] has a relatively limited range of variation within MCs, so we can better test its correlations (\fig{correl_nt}).      
            
\section{An Estimation of the Flux Rope Axis Length} 
      \label{sect_Length}      

 Since we found no significant dependences along the FR axis of $\Bo$, $\Ro$ and $\nt$ in the previous section, we simply need the FR length to estimate the total magnetic helicity of MCs. At 1~AU, this length was typically taken in the range $[0.5,2.5]$~AU in previous studies (\sect{Hestimation}). In this section, the length is estimated from the information derived statistically on large samples of MCs.
 
\subsection{Method to Derive a Mean Axis Shape} 
      \label{sect_Method_Axis}

With a set of MCs, one can define an observed probability distribution for each parameter of the fitted model.  \citet{Janvier13} have developed a method to deduce a generic MC axis shape from the observed probability $\pobsl$.  The main idea is that MCs are crossed at various locations, \ie\ at different $\lA$ values along their axis.  Then, the observed probability $\pobsl$ is a consequence of the axis shape, with more detections expected as the local orientation of MC axis is further away from the radial direction from the Sun (\fig{schema}). 
The statistical analysis supposes that all MCs have a comparable axis shape with only a scaling factor in the angular extension ($2 \phimax$, \fig{schema}). Indeed, the probability $\pobsl$ was shown to be nearly independent of the MC parameters such as field strength, radius and inclination on the ecliptic when the MCs were analysed in sub-groups \citep{Janvier13}.  Further they showed with a synthetic MC axis model that the angular extension $2 \phimax$ has almost no effect on $\pobsl$.  This justifies the analysis of all the MC set together, and the derivation of a mean axis shape from $\pobsl$ with $\phimax$ as the only free parameter.

Below, we first shortly summarise the analysis of \citet{Janvier13} before extending it to derive the curvilinear abscissa along the axis, and then its length.
The MC axis is supposed to be inside a plane and it is described with cylindrical coordinates [$\rho, \varphi$] (\fig{schema}).  The probability of crossing a MC can be expressed either in function of $\varphi$ as $\pvphi (\varphi) \rmd \varphi$, or in function of $\lA$ as $\pobsl ~\rmd \lA$ and these two probabilities are equal. 
Since CMEs are launched from a broad range of solar latitude and any longitude over the time scale of the analysed MC set (almost a solar cycle), the MCs are expected to be crossed with a uniform distribution in $\varphi$, so $\pvphi = 1/(2\phimax)$ with the normalisation of the probability to unity.
At the difference of \citet{Janvier13}, here we do not symmetrise $\pvphi (\varphi )$ and $\pobsl$, so we keep separated positive and negative values of $\varphi$ and $\lA$.
The above equality of probabilities implies   
  \BE \label{eq_dvarphi}
  \rmd \varphi = 2 ~\phimax ~\pobsl ~\rmd \lA  \, .
  \EE
Its integration defines $\varphi$ as a function of $\lA$ as 
  \BE \label{eq_varphi}
  \varphi (\lA, \phimax ) = 2 ~\phimax \int_{0}^{\lA} \pobs (\lA') ~\rmd \lA'  \, .
  \EE
Next, we relate $\rho$ to $\lA$ by expressing $\lA$ as the angle between the radial direction [$\ur$] and the normal to the axis [$\uvec{n}$] (\fig{schema}), which writes as  
  \BE \label{eq_tanLambda}
  \rmd \ln \rho = - \tan (\lA ) ~\rmd \varphi \, .
  \EE
Using \eq{dvarphi},  the integration of \eq{tanLambda} implies
  \BE \label{eq_rho}
  \ln \rho (\lA, \phimax ) = -2 ~\phimax  
            \int_{0}^{\lA} \tan (\lA') ~\pobs (\lA') ~\rmd \lA'
           +\ln \rho_{\rm max}  \, .
  \EE
\eqs{varphi}{rho} define a generic flux rope shape as a parametric curve ($\rho (\lA), \varphi (\lA)$) in cylindrical coordinates, from the  probability distribution $\pobsl$ derived from the analysis of a MC set.

We extend the previous analysis by defining the curvilinear elementary length [$\rmd \sa$] along the axis as
  \BE \label{eq_ds}
  \rmd \sa = \sqrt{ (\rmd \rho)^2 + (\rho ~\rmd \varphi)^2} 
           = \sqrt{1+ \left( \frac{\rmd \ln \rho}{\rmd \varphi} \right)^2} 
                  ~\rho ~\rmd \varphi 
           =  \frac{\rho ~\rmd \varphi}{\cos (\lA)}  \, ,
  \EE
  after introducing \eq{tanLambda}.
Proceeding as above for the derivation of $\varphi (\lA)$ and $\rho (\lA)$, the curvilinear abscissa with origin at the apex is
  \BE \label{eq_s}
  \sa (\lA, \phimax ) = 2 ~\phimax  
            \int_{0}^{\lA} \frac{\rho (\lA') ~\pobs (\lA') }{\cos (\lA')}
                            ~\rmd \lA' \, .
  \EE

\begin{figure*}  
 \centerline{\includegraphics[width=0.7\textwidth]{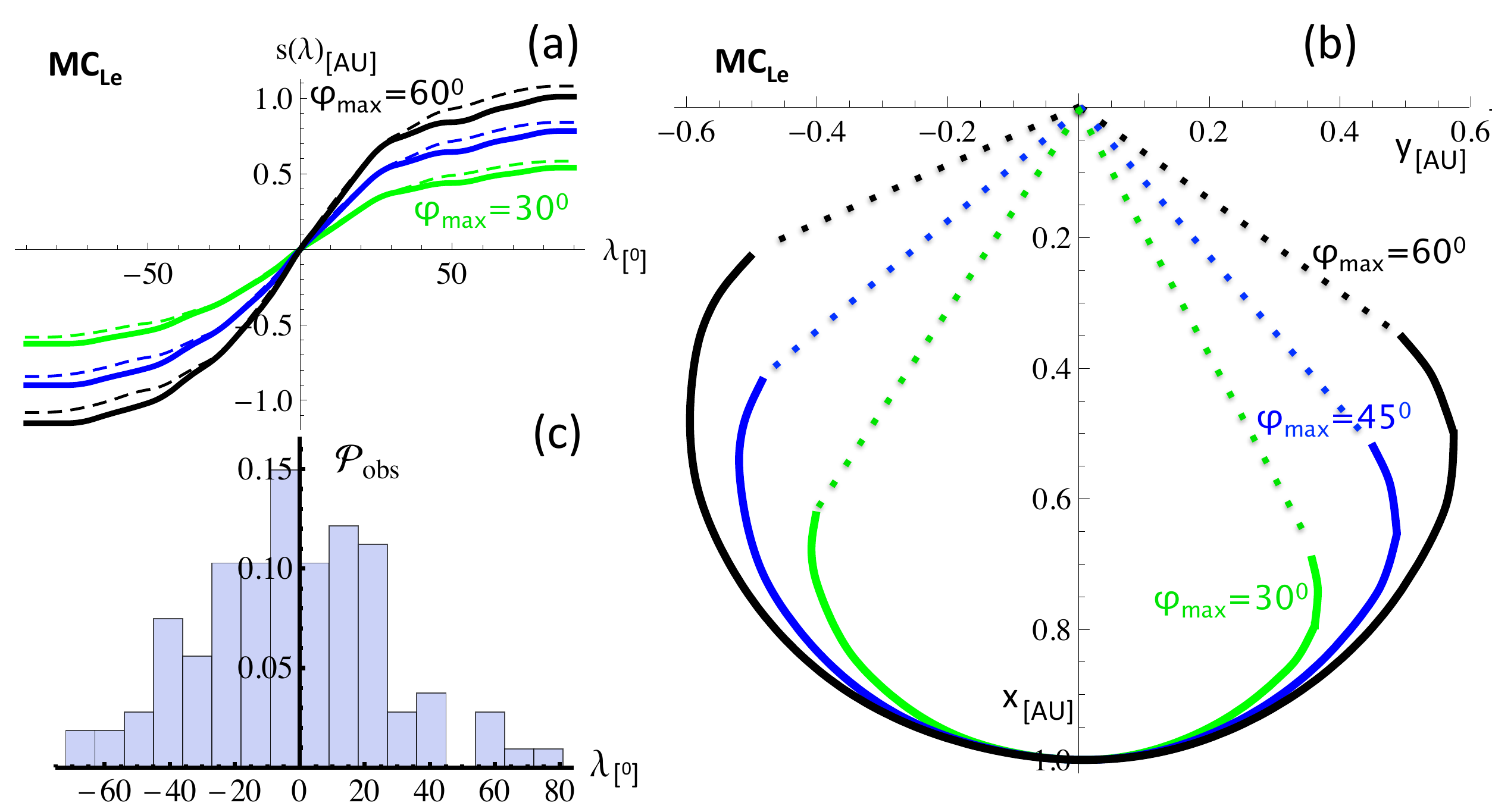} }
\caption{Effect of the angular extension $\phimax $ (defined in \fig{schema}) on (a) the curvilinear abscissa [$\sa$ in AU] along the FR axis  and (b) the shape of the FR axis for \MCb\ set. 
All the curves are derived from the probability distribution [$\pobsl$] of \MCb\ set shown in panel (c). The dashed-lines in panel (a) represent the case where $\pobsl$ is set to be symmetric in $\lA$.  This forced symmetry has a small effect on $\sa (\lA )$. 
The dotted lines in panel (b) represent the flux rope legs extrapolated to the Sun by a radial segment (used to computed $\Ltot$ in \eq{Ltot}).
}
 \label{fig_axis_30-60_Lepping}
\end{figure*}

\begin{figure*}  
 \centerline{\includegraphics[width=0.7\textwidth]{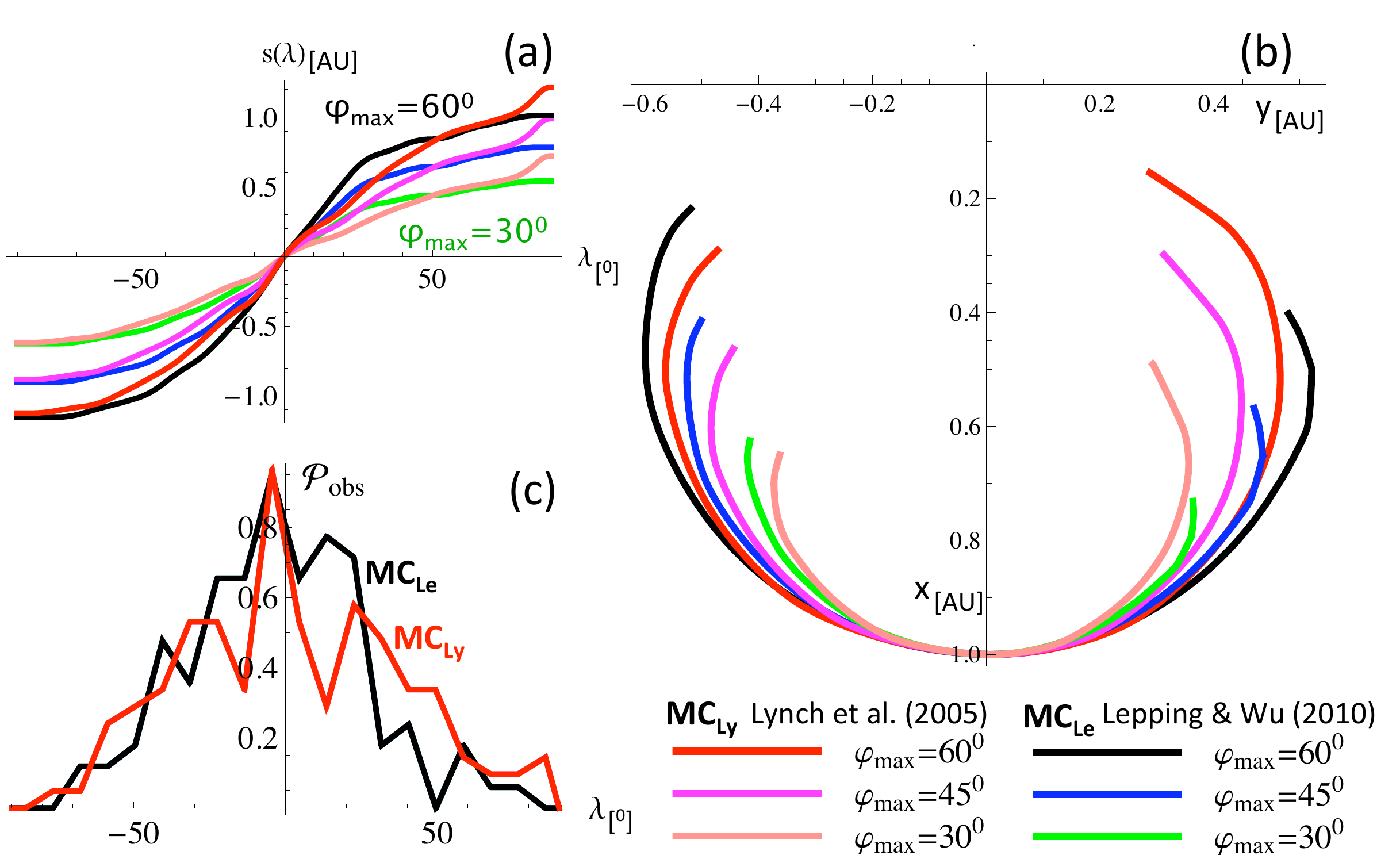} }
\caption{Comparison of the results between the two analysed MC sets.
The results are shown for three values of the parameter $\phimax $. (a) The curvilinear abscissa [$\sa$ in AU] along the FR axis (defined in \fig{schema}) and (b) the shape of the deduced MC axis.  The curves are derived from the probability distributions [$\pobsl$] shown in panel (c) with curves, rather than histogram, for comparison.  The black distribution of \MCb\ set is the same as in \fig{axis_30-60_Lepping}c and corresponds to the black, blue and green curves in panels (a) and (b).  The red distribution is derived for \MCa\ set and corresponds to the red, magenta and pink curves.
}
 \label{fig_axis_30-60_Lepping_Lynch}
\end{figure*}

At the limit $|\lA'| \rightarrow 90\degree$, $\cos (\lA') \rightarrow 0$ at the denominator. However, in the observations $\pobs (\lA')$ strongly decreases with $|\lA'|$ and vanishes above $|\lA'|> 80 \degree$ (Figures~\ref{fig_axis_30-60_Lepping}c and \ref{fig_axis_30-60_Lepping_Lynch}c) so that the integral is not singular, but rather $\sa (\lA)$ is typically flat for large $|\lA|$ values (Figures~\ref{fig_axis_30-60_Lepping}a and \ref{fig_axis_30-60_Lepping_Lynch}a).   

\begin{figure*}  
 \centerline{\includegraphics[width=0.35\textwidth]{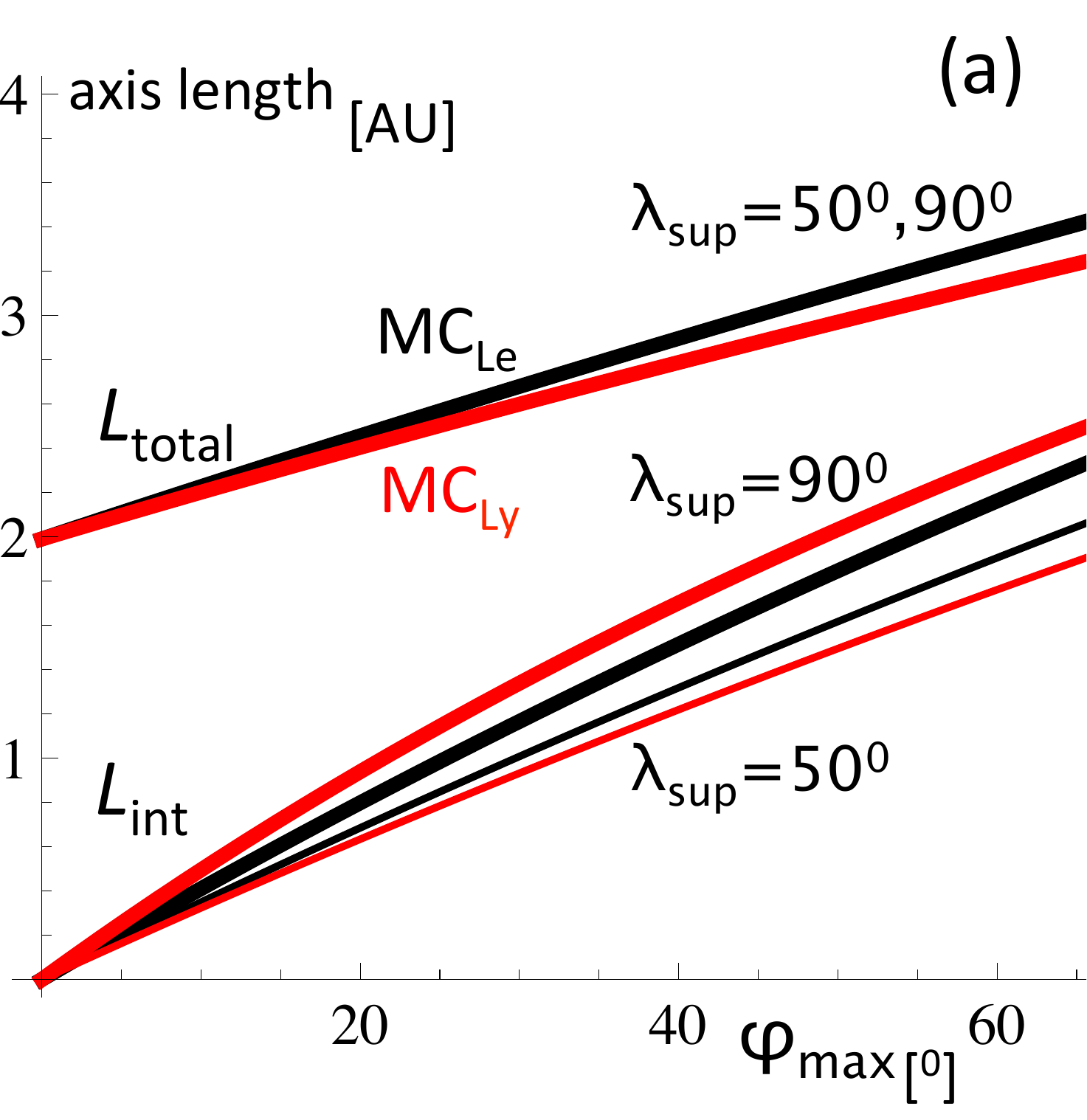}
             \includegraphics[width=0.35\textwidth]{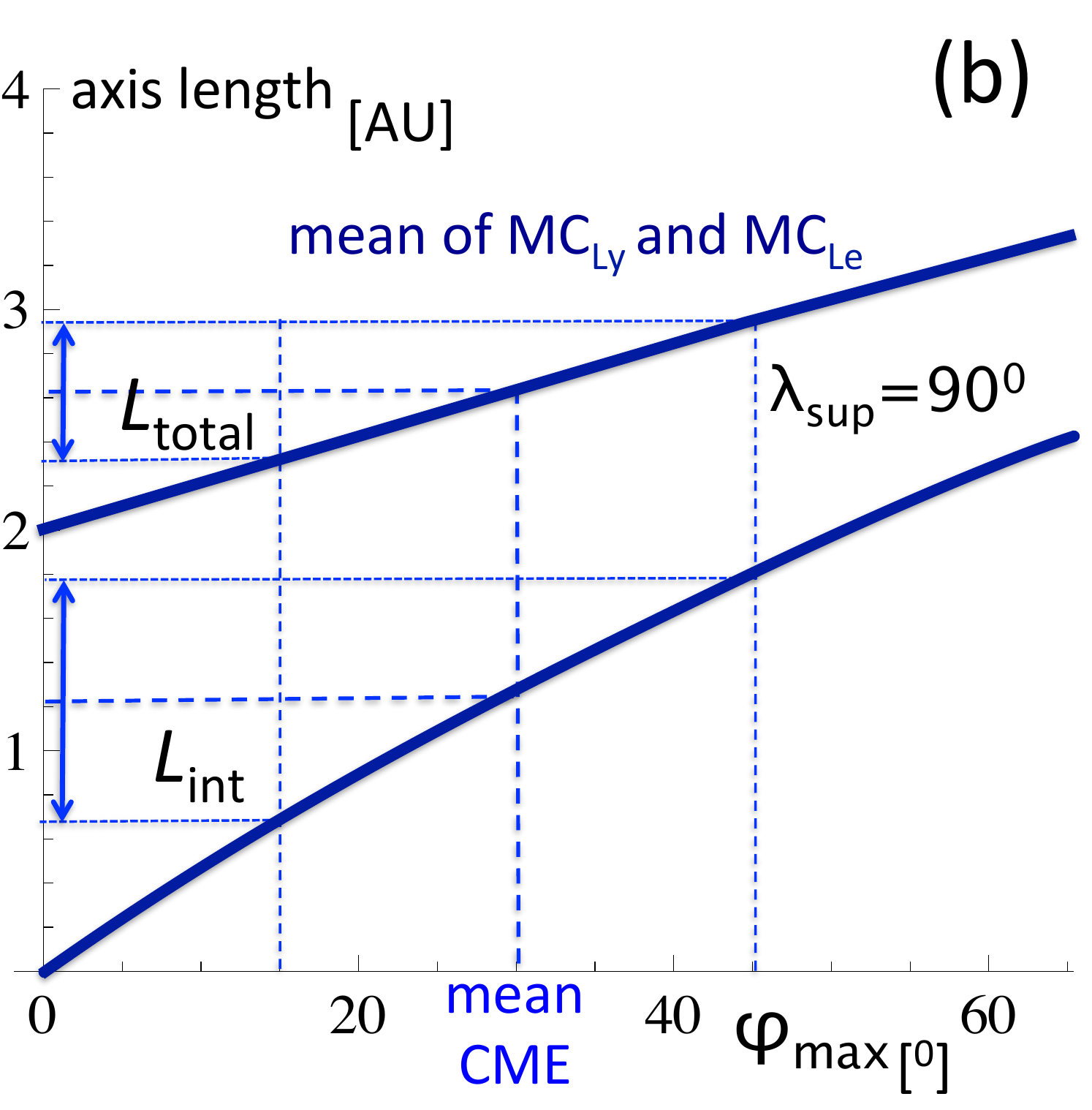} }
\caption{Dependence of the lengths $\Lint$, \eq{Lint}, and $\Ltot$, \eq{Ltot}, with the angular extension $\phimax$.   The FR has its apex located at 1~AU and is still attached at the Sun to compute $\Ltot$.   
(a) Effect of the integration upper limit $\lS$ for both \MCa\ and \MCb\ sets (in red and black, respectively).
(b) The mean value and the typical range of $\phimax$ derived from limb CMEs observed with coronagraphs (Wang \etal\ 2011) are used to estimate a mean value and a range of $\Lint$ and $\Ltot$ (blue arrows) from the mean results of \MCa\ and \MCb\ (blue curves).
}
 \label{fig_length}
\end{figure*}  

\subsection{Mean Axis Length} 
      \label{sect_Mean_Length}
       
The application of \eqs{varphi}{rho} to the \MCb\ set is shown in \fig{axis_30-60_Lepping}b for three values of $\phimax $.  The deduced axis shape is only weakly asymmetrical between the two sides, then comparable to the schema drawn in Figure 2 of \citet{Burlaga98}.  This result is also comparable to Figure 2 of \citet{Zurbuchen06} for the front part while the legs are bended with a Parker-like spiral.   The spiral shape is due to the rotation of the Sun carrying the anchored field lines.  

The curvilinear abscissa is evolving more around the apex ($\lA=0$) than at larger $|\lA|$ values (Figures~\ref{fig_axis_30-60_Lepping}a and \ref{fig_axis_30-60_Lepping_Lynch}a). There is also a weak asymmetry between the legs as shown by comparing the continuous curves with the dashed curved showing $s(\lA)$ computed with a symmetric probability (imposing $\pobs (\lA)=\pobs (-\lA)$).   As expected, the angular extension $\phimax $ has a significant effect on the curvilinear abscissa (\fig{axis_30-60_Lepping}a).

The results with the \MCa\ set are close to the ones for \MCb\ set  with the minor difference of $s(\lA)$ increasing more sharply close to $\lA=+80\degree$ for \MCa\ set (\fig{axis_30-60_Lepping_Lynch}a).
This is a consequence of the local maximum in the positive tail of $\pobs (\lA)$ (red curve in \fig{axis_30-60_Lepping_Lynch}c). It has also implications for the axis shape with a positive $\lA$ leg extending more towards the Sun for \MCa\ than \MCb\ set (\fig{axis_30-60_Lepping_Lynch}b).  The other local peaks or dips in $\pobs (\lA)$ (\fig{axis_30-60_Lepping_Lynch}c) have only a weak effect on $s(\lA )$ (\fig{axis_30-60_Lepping_Lynch}a).  This is a consequence of the integration averaging effect.
 
The quantities $\Bo$, $\Ro$ and $\nt$ are statistically independent of $\lA$ in the range $[-50\degree, 50\degree]$ (\sect{Physical}). Then, a minimal length to compute $H$ is within this $\lA$ interval (unless the flux rope reconnects with the solar wind magnetic field).
Another estimate of the length is to extend this interval to the full range of $\lA$: $[-90\degree, 90\degree]$. More generally, the length can be estimated for the range $[-\lS , \lS ]$ as
  \BE \label{eq_Lint}
  \Lint  (\lS, \phimax ) = \sa (\lS, \phimax ) -\sa (-\lS, \phimax )\, .
  \EE
\fig{length}a shows the evolution of $\Lint  (\lS, \phimax )$ in function of $\phimax $. It is nearly a linear function of $\phimax $, because the curvilinear abscissa $\sa (\lS, \phimax )$ is proportional to $\phimax $ (\eq{s}).  
However, the linearity is only approximative because an extra dependence on $\phimax $ is present in $\rho (\lA')$ (\eq{rho}). This dependence is weaker since $\ln \rho $, and not $\rho $, is a linear function of $\phimax $.  
Next, there is only a slight increase of $\Lint$ between $\lS = 50\degree$ and $\lS = 90\degree$ (thin and thick lines, respectively, in \fig{length}a). 
The lengths computed with the two different sets \MCa\ and \MCb\ are comparable (red and black continuous lines, respectively). The slightly larger $\Lint$ for $\lS = 90\degree$ and \MCa\ set is a consequence of the larger extension of the computed axis towards the Sun (\fig{axis_30-60_Lepping_Lynch}b). This effect is reversed for $\lS = 50\degree$.    
  Finally, for $\phimax $ around $30\degree$, the typical CME extension observed with imagers (see below), $\Lint$ could simply be approximated by the linear function $0.2+3.2~\phimax/90$.  
  
If the flux rope is still attached to the Sun by both legs, a lower estimate of the total length is given by adding radial straight lines linking the photosphere to the ends of the axis shape found above (\figs{axis_30-60_Lepping}{axis_30-60_Lepping_Lynch}).  Since the Parker spiral is close to the radial direction close to the Sun, this straight line approximation provides only a slight underestimation of the length.  This total length writes
  \BA 
  \Ltot (\lS, \phimax ) &=& \Lint (\lS, \phimax ) \nonumber \\
  &+& \rho (\lS, \phimax ) + \rho  (-\lS, \phimax ) -2 R_{\odot} \, . \label{eq_Ltot}
  \EA
where $R_{\odot}$ is the Sun radius. $\Ltot$ is even closer to a linear function than $\Lint$
(\fig{length}a), because the contribution of the straight leg parts ($\rho (\pm \lS, \phimax ) - R_{\odot}$) nearly compensates the contribution of $\Lint$ with increasing $\phimax$ values.  The same is true for the dependence with $\lS$: $\Ltot (\lS, \phimax )$ curves superpose each other very well in \fig{length}a for $\lS = 50 \degree$ and $90 \degree$.  Finally, the linear function $2+1.8~\phimax /90$ approximates $\Ltot $ very well. 

An estimation of $\phimax$ with \insitu\ measurements is generally not possible because only a few MCs are crossed by several spacecraft \citep[\eg , ][]{Burlaga90,Ruffenach12}.  However, its estimation could be given from CME-imaging, which records a higher number of CMEs.
Observations of CMEs situated close to the Sun and at the limb minimise the projection effects, although the tilt of the flux rope axis cannot be inferred. Since the orientation of the flux rope is not determined, this supposes a comparable angular extension of CMEs along and across the flux rope.  The typical nearly circular observed shape of CMEs directed toward the observer (full halo CMEs) justifies this hypothesis. 
\citet{Wang11} derived from the Large Angle and Spectrometric Coronagraph (LASCO) a mean $\phimax$ of $30 \degree$ for limb CMEs and for 65\% of the CMEs, its values lie within the interval $[15 \degree , 45  \degree ]$. Since these values are deduced from coronagraphs, which image the densest parts of the CMEs, namely the sheath region preceded by a shock, the intervals given for $\phimax$ are not strictly speaking those of the MC axis.  Indeed, the above values are slightly too large, for example in a well observed case, $\phimax$ is about $10\degree$ larger for the MC sheath than for the FR axis \citep{Janvier13}.  However, since the MC axis extension angles are not generally known, we report the $\phimax$ values estimated from CMEs in \fig{length}b to derive $\Lint = 1.3 \pm 0.6$~AU and $\Ltot = 2.6 \pm 0.3$~AU, then $\Ltot \approx 2 \Lint$. 
 

\section{Total Amount of Global Flux Rope Quantities} 
      \label{sect_Total_Amount}     

\subsection{From Local to Total Estimations} 
      \label{sect_Local_Total} 

   The above results of \sectss{Observations}{Length} are applied to compute the magnetic axial flux $F$ and helicity $H$ for each FR detected at 1~AU.
The results are summarised in the distribution functions $\dQsdRobs{F}$ and $\dQsdRobs{H}$, dependent of $R$. They provide the amount of magnetic flux and helicity per unit radius and time.   These distributions are related to the distribution of FR number, $\dQsdRobs{N}$, as $\dQsdRobs{Q}=Q~ \dQsdRobs{N}$ with $Q=F$ or $H$.  
   
The distribution $\dQsdRobs{Q}$ measures the local distribution of $Q$ as estimated by the spacecraft over an interval of time.  We are also interested on the total amount of these quantities crossing the sphere of radius $D=1$~AU.
In order to convert the local distribution to a global distribution of FRs travelling at least up to 1~AU, \citet{Janvier14} have estimated the probability to detect a FR on the sphere of radius $D$ supposing a uniform distribution of FRs in longitude and in a latitude band $\pm \latmax$. This portion of sphere has a surface $S_{\rm sp} = 4\pi \sin \latmax D^2$.
The FR extension projected on the sphere, so its apparent surface $S$, is estimated by $S=2 R L_{\rm p}$ where $L_{\rm p}$ is the FR axial extension projected radially on the sphere. The probability to detect this FR is $P_{\rm FR} = S/S_{\rm sp}$.  Then, the total distribution function is $\dQsdRtot{Q} = \dQsdRobs{Q} \times 1/ P_{\rm FR}$.
This computation corrects the local spacecraft measurements both to estimate the total number of FRs launched from the Sun and to take into account the lower probability to detect a FR with a lower radius (as its cross section viewed by the spacecraft is lower).  
Finally, all the distributions are averaged over the time period of the \insitu\ observations and are computed per year to be compared. 

 The projected length $L_{\rm p}$ is estimated from the mean angular extension of CMEs close to the solar limb: $\phimax = 30 \pm 15 \degree$ \citep{Wang11} providing $L_{\rm p}\approx 1 \pm 0.5$~AU. 
From the latitude distribution of the expected solar sources, \citet{Janvier14} selected $\latmax=45 \degree$.  The global distribution would be simply multiplied by a factor 0.8 (resp. 1.4) if $\latmax=30 \degree$ (resp. $60 \degree$) would be used instead.  The results of the above procedure and the value $\latmax=45 \degree$ were back up by checking that the computed total number of MCs from \insitu\ data matches the expected total number of MCs derived from coronagraph observations of CMEs (see \sect{Cumulative}).

 We also study the content of helicity in small FRs in the solar wind. 
We analyse a sample of 125 small flux rope events presented by \cite{Feng07}, and also its extension made in \citet{Feng08} \citep[see][for details about several features of these two samples]{Janvier14c}. 
The same procedure is applied to small FRs and MCs while it is not known whether small FRs are that broadly extended along their axis than MCs.  Then, we simply use the same $\phimax$ and  $L_{\rm p}$ values for small FRs than for MCs.
This choice has in fact a negligible effect on the total magnetic flux and helicity estimations (\sect{Cumulative}), and the small FR contribution is even expected to be smaller as small FRs are likely to be coherent flux ropes only on length scales smaller than $L_{\rm p}$ used here.

\begin{figure*}  
 \centerline{ \includegraphics[width=0.7\textwidth]{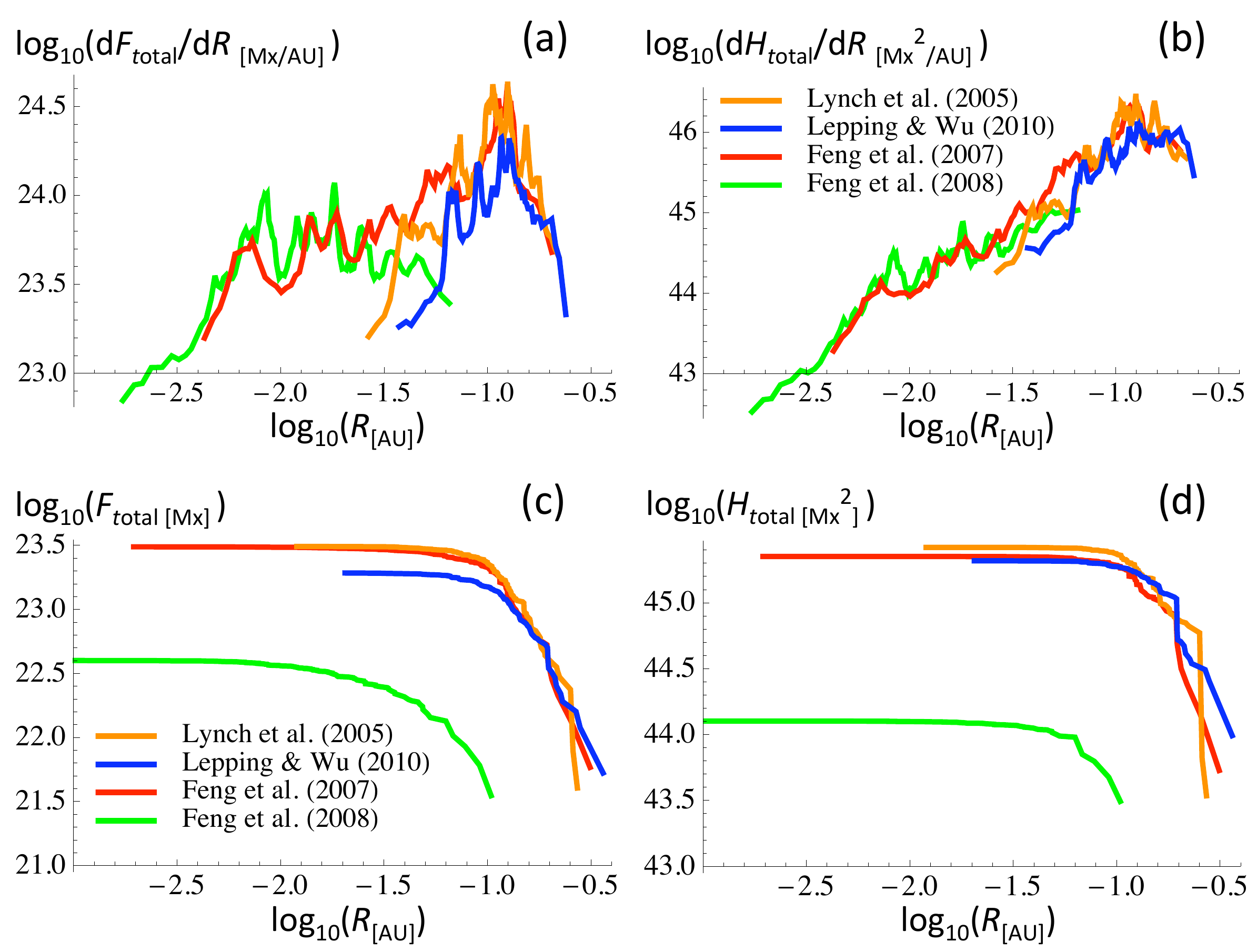} }
\caption{(a,b) Distributions of magnetic axial flux and helicity for the four studied lists of MCs and small FRs. $\rmd Q$ is the amount of $Q$ ($=F$ or $H$) in the radius interval $\rmd R$ per year.  
         (c,d) Cumulative distribution functions of $F$ and $H$ 
computed per year with an average over the time period of the lists (Table~\ref{Table_Max}).  The number of events in all curves are corrected from the apparent FR cross section projected on the sphere with 1~AU radius (\sect{Local_Total}) and the summations are done from the larger to the smaller radius (\eq{Qtotal(R)}). $H$ is computed with $\Ltot = 2.6$~AU (\sect{Mean_Length}).
}
 \label{fig_cumul}
\end{figure*}  

\subsection{Distribution Functions} 
      \label{sect_Distribution} 

The distribution functions can be estimated with histograms. However, a uniform binning of $R$ is not suited due its large range of variation and the large variation of the number of FRs per bin.  In fact the bin size should be adapted to the number of FRs detected in each range of $R$. Then, \citet{Janvier14} developed a technique, called the partition method, where the bin widths are computed to have the same number, $\Npart$, of FRs in each bin in order to have a uniform statistical noise across the bins.  The data are first ordered by growing value of $R$.  The binning with $\Npart$ flux ropes is computed starting from the lower $R$ values, then  progressively shifting upward the bin window by one FR.  This provides a smoothing of the fluctuations over $\Npart$ flux ropes. We use $\Npart=10$ as a compromise between decreasing the fluctuations and resolving the variations of the distributions.   

\fig{cumul}a shows that $\dQsdRtot{F}$ of MCs dominates $\dQsdRtot{F}$ of small FRs, but only by a factor $\approx 3$.
The MC contribution is peaked while the small FR contribution is almost independent of $R$ \citep[for $\log_{10} R < -2.3$, the decrease is due to a selection effect on small FR orientation, see][]{Janvier14}. 

Both distributions, $\dQsdRtot{F}$ and $\dQsdRtot{H}$, are maximum for $R\approx 0.13$ AU with the difference that $\dQsdRtot{F}$ is peaked around this maximum while $\dQsdRtot{H}$ is nearly flat in the range $0.08 \leq R \leq 0.2$~AU (\fig{cumul}a,b).  For $R \leq 0.06$~AU, $\dQsdRtot{H}$ is already one order of magnitude below its plateau value so that the MC contribution strongly dominates for helicity. 
Finally, the slightly lower distributions for \citet{Lepping10}, compared to the ones computed from the two other lists having MCs \citep{Lynch05,Feng07} are due to a more severe selection of MCs so a lower number of detected MCs by a factor $\approx 2$. 

The distribution $\dQsdRtot{N}$ is a strongly decreasing function of $R$ proportional to  
$R^{-2.4}$ in the range of small FRs \citep[see Figure 3 of ][]{Janvier14}.  
A bump is present in this distribution for MCs, still the number of FRs is dominated by the small FRs (their Figure 6).  By contrast, $\dQsdRtot{F}$ and even more $\dQsdRtot{H}$ distributions are larger in the MC region (\fig{cumul}a,b).  
The smaller FR are not enough numerous, \ie\ $\dQsdRtot{N}$ is not steep enough, to balance the $R^2$ and $R^3$ factors and the less variable factors $\Bo$ and $\Bo^2$ present in \eqs{FzL}{HL}  \citep[$\Bo$ is on average an increasing function of $\Ro$, ][Figure~3]{Janvier14c}.

Interestingly, the distribution $\dQsdRtot{H}$ \p{is close to a power law of the radius with an exponent $\approx 2$.  The dominance of the large scales may be a consequence of the inverse cascade of helicity, as found in MHD studies where helicity was found to be transferred} from small scales to larger scales \citep[][and references therein]{Alexakis06}. \p{This property could be a consequence of the MHD evolution of the corona and even of the solar dynamo which build the coronal magnetic field (while flux ropes are mostly transported in the interplanetary medium with only some erosion).}
 \p{The plasma composition is in favor of a coronal formation \citep{Feng15}, and other characteristics of small FRs point toward different formation mechanisms 
than MCs (see Section 5.2 in \citealt{Janvier14c}). Then, one mechanism} (\eg\ the tearing instability) cannot be put forward to explain this distribution of both small FRs and MCs. 
Since they also have similar characteristics at 1~AU, both the formation and the propagation process from an early stage in the corona to the interplanetary medium might play a role in 
transferring magnetic helicity from smaller scales to larger ones. We may find an answer to this question with a future study on the helicity partition at solar distances $<$1~AU.

\subsection{Cumulative Functions} 
      \label{sect_Cumulative} 

We are also interested on the total amount of magnetic flux and helicity crossing the sphere of radius 1~AU per unit time, in order to have a global budget of these quantities launched by the Sun.

   We define similarly as above a global quantity by $Q$.  We use below $Q=1,F,H$ to compute the total number, the magnetic flux and helicity of FRs.  The amount of $\rmd Q_{\rm total}$ in the range of radius $\rmd R$ is $Q~(\dQsdRtot{N}) ~\rmd R$ and we define the cumulative function of $Q$ starting from the largest FRs since they are less numerous:
  \begin{equation}  \label{eq_Qtotal(R)}
  Q_{\rm total}(R) = \int_{R}^{R_{\rm max}} \frac{\rmd Q_{\rm total}}{\rmd R}(R')  ~\rmd R' 
                   = \int_{R}^{R_{\rm max}} Q(R') ~\frac{\rmd N_{\rm total}}{\rmd R}(R') ~\rmd R'  \,.
  \end{equation}
$Q_{\rm total}(R)$ can be computed by integrating the distributions shown in \fig{cumul}a,\-b, or simpler by a summation of the contribution of each MC starting from the largest ones and applying the conversion factor associated to the probability to detect this FR ($P_{\rm FR}$, \sect{Local_Total}). Indeed, the fluctuations are naturally averaged in a cumulative function so we show in \fig{cumul}c,d the results of this second approach which is more direct and simpler.

 The computed total number of MCs match approximately the number of MCs expected from the number of observed CMEs \citep{Janvier14}. More precisely, the counts derived from \citet{Lynch05} and some CME catalogs are close to each other, while the same pairing is true between the counts derived from \citet{Lepping10} and some other CME catalogs (see their Figure 6).  The difference of counts between the pairs is about a factor 2 (Table~\ref{Table_Max}) which is linked to the slightly different criterium used to defined both MCs and CMEs.  This correspondence backups the above procedure which transforms the local measurements of Wind or ACE to global estimations.   It shows also that the small FRs, much more numerous than MCs by at least a factor 10 (Table~\ref{Table_Max}), are not associated to CMEs.
    
While the small FRs dominate by their number, they provide a small contribution to $F_{\rm total}$ and $H_{\rm total}$ (\fig{cumul}c,d and Table~\ref{Table_Max}).  The small FRs from the list of \citet{Feng08} provide a factor 8 lower magnetic flux and a factor 20 less magnetic helicity per year than the MCs of the list of \citet{Lynch05}.  Both cumulative curves are indeed flat below some $\Ro$ value.  The main contribution to the cumulative curves is from the larger MCs as expected \citep{Lynch05}.   Half of the contribution for the axial magnetic flux is from MCs with a radius $\Ro$ larger than $\approx 0.13$~AU (with only slight variations between the three first lists which contain MCs,  Table~\ref{Table_halfMax}). The contribution for magnetic helicity is even from larger scale MCs ($R \gtrsim 0.14$ to $0.2$~AU).

The results from the three lists containing MCs are the closest for $H$ with $\Htot= 2.3 \times 10^{45}$~Mx$^2$ and 
at most a variation of 13\% between lists compared to a factor 2 with the number of MCs. Indeed, the identification and modeling of large MCs, 
which define dominantly $\Htot$, is easier than smaller ones which typically have properties less contrasted, when they are compared to the surrounding solar-wind properties.

\setlength{\tabcolsep}{6pt}   
\begin{table*}   
\caption{Maximum values of the cumulative function of $\Ntot$, $\Ftot$ and $\Htot$ estimated during one year.  They are averages over the time period indicated in the first column. $H$ is computed with $\Ltot = 2.6$~AU (\sect{Mean_Length}). The presence of small FRs is indicated by sFR and of magnetic clouds by MC in the third column. 
}
\label{Table_Max}   
\begin{tabular}{c@{~}ccccc}     
  \hline                  
time period & reference of FR list   & type   &$\Ntot$&$\Ftot$&$\Htot$  \\
            &                        &        &number & $10^{22}$ Mx & $10^{42}$ Mx$^2$\\
  \hline 
1995-2003   & \citet{Lynch05}   &  MC    &~\,850 &   31  & 2600 \\
1995-2009   & \citet{Lepping10} &  MC    &~\,390 &   19  & 2100 \\
1995-2001   & \citet{Feng07}    &MC$+$sFR&  4600 &   31  & 2200 \\
1995-2005   & \citet{Feng08}    &  sFR   & 7040 & ~\,4  &~~130 \\
  \hline
\end{tabular}

\end{table*}   

\begin{table*}   
\caption{FR radius [$\Ro_{\rm h}$ in AU] where the cumulative function of $\Ntot$, $\Ftot$ and $\Htot$ (shown in \fig{cumul}) reach half their maximum value (given in Table~\ref{Table_Max}). The presence of small FRs is indicated by sFR and of magnetic clouds by MC in the third column.   
}
\label{Table_halfMax}   
\begin{tabular}{cccccc}     
  \hline                  
time period & reference of FR list & type   & $\Ro_{\rm h,N}$ & $\Ro_{\rm h,F}$ & $\Ro_{\rm h,H}$ \\
  \hline 
1995-2003 & \citet{Lynch05}   &  MC    & 0.066 &  0.13 & 0.15 \\
1995-2009 & \citet{Lepping10} &  MC    & 0.084 &  0.14 & 0.20 \\
1995-2001 & \citet{Feng07}    &MC$+$sFR& 0.008 &  0.12 & 0.14 \\
1995-2005 & \citet{Feng08}    &  sFR   & 0.005 &  0.04 & 0.07 \\
  \hline
\end{tabular}
\end{table*}   

\subsection{Total Magnetic Flux and Helicity from MCs and CMEs} 
      \label{sect_Total-F-H} 

The amount of magnetic flux carried by MCs in one year is important (Table~\ref{Table_Max}).  With a typical  large magnetic flux of $10^{22}$~Mx, but not exceptional for an active region (AR), this implies that all the MCs launched from the Sun carry on average the magnetic flux of 20 to 30 ARs in one year.  The amount of launched unsigned magnetic helicity is even more important, as follows. The magnetic helicity injected at the photospheric level during the full emerging phase of an AR is typically $10^{43}$~Mx$^{2}$ for a flux of $10^{22}$~Mx \citep[\eg , ][]{Jeong07,Lim07,Tian08,Jing12}.   This implies that MCs carry per year the magnetic helicity of about 200 emerging ARs.   

The above estimations of $\Ftot$ and $\Htot$ are computed only for the estimated total number of MCs crossing 1~AU in one year.  However, MCs are detected in about only one third of ICMEs on average over the solar cycle \citep[\eg , ][and references therein]{Richardson10}. In about one third of ICMEs, a magnetic field rotation is detected but it is not coherent enough or a low enough proton temperature is not found, so they are not classified as MCs but as cloud-like events \citep{Lepping05}.
If the lower detection rate of MCs is simply due to the spacecraft passing on the side or missing the flux rope, as advocated by \citet{Jian06}, so if all ICMEs have a flux rope inside \citep[as recent results point out: ][]{Gopalswamy13,Makela13}, the amounts of $\Ftot$ and $\Htot$ shown in \fig{cumul} and Table~\ref{Table_Max} need to be multiplied by a factor $\approx 3$.

On the contrary, the amount of estimated helicity could be lower because some MCs are likely no longer attached to the Sun when observed at 1~AU.   Not taking into account the length of MC legs decreases the length and the helicity estimated by a factor 2 (\sect{Mean_Length}).  
   Next, the erosion of the FR by reconnection with the solar-wind magnetic field is not taken into account in the above lists and deduced results.   The analysed MCs contain the FR remaining intact at 1~AU, and a part of the reconnected flux \citep[called a back region by][]{Dasso06}.  Then, the above flux and helicity estimates are in between the ones from the remaining FRs at 1~AU and the FRs launched from the Sun. Since the average amount of reconnected flux is large, about 40 \% of the total azimuthal 
magnetic flux, the above helicity estimate is expected to be a factor around 2 too low for the helicity launched by the Sun.   
   Finally, the FR cross section is typically flat by a factor 2 to 3 on average \citep{Demoulin13}.  Compared to the cylindrical model used to fit the data, this introduces an underestimation of the helicity by a factor slightly below 2 to 3 \citep[helicity is near proportional to the aspect ratio for $b/a \geq 2$, see Figure 8a of][]{Demoulin09b}.   To summarise, the over estimation of helicity, implied by supposing all MCs observed at 1 AU as attached to the Sun, is likely to be over-compensated by the other factors described above.  Then, our estimates of magnetic helicity (\fig{cumul}, Tables~\ref{Table_Max},\ref{Table_halfMax}) are expected to be underestimated by at least a factor 2, and plausibly a factor 6 if non-MC CMEs carry the same amount of helicity than MCs.

\subsection{Helicity Budget over a Solar Cycle} 
      \label{sect_Cycle} 

Below we estimate the total unsigned helicity that leaves the Sun over a solar cycle. The list of \citet{Lepping10} covers most of a solar cycle 23. While the two other lists are more restricted in time, Table~\ref{Table_Max}, 
they provide comparable values of helicity transported per year. Then, we convert these results to the total amount of unsigned magnetic helicity transported by MCs during 
one solar cycle by both hemisphere (assuming a constant mean helicity per year): $\Hmc \approx 2.5 \times 10^{46}$ Mx$^2$.  
This is one order of magnitude larger than the estimation of \citet{Bieber95} from solar cycles 20-22, and a factor 2.5 larger than the one of \citet{DeVore00} from solar cycle 21.  However, in this last case the difference is mostly due to difference of axis length used: $0.5$~AU for \citet{DeVore00} compare to $2.6$~AU here (\sect{Mean_Length}).  Then, within a factor 2, our results agree with the order of magnitude estimated by \citet{DeVore00}.

Before comparing our result to the solar source estimate, it is worth to compare it to other solar phenomena to appreciate its magnitude.
$\Hmc$ is three orders of magnitude larger than the total helicity injected in the quiet Sun \citep{Welsch03} and one order of magnitude lower than the total helicity injected in open field of coronal holes \citep{Berger00} while the total unsigned magnetic flux involved have comparable magnitudes. 
Next, the analytical expression for the magnetic helicity contained in a simplified Parkerian solar wind for a period of a solar rotation was computed by \cite{Bieber87} (see their Equation (8)).
From this expression, an helicity near $7 \times 10^{47}$ Mx$^2$ is obtained for a complete solar cycle.  It is a value comparable to the solar estimation of \citet{Berger00} with a constant open flux $\approx 4 \times 10^{22}$ Mx per magnetic polarities.  Then, the ejection of FRs from the Sun is an efficient mechanism to eject magnetic helicity, but less efficient than the direct solar rotation twisting the open flux.
 
Next, the MCs are not related to the quiet Sun nor to the solar open flux, but to the solar dynamo and ARs, as follows.  Differential rotation in the convection zone creates an opposite amount of magnetic helicity in each solar hemisphere.  The amount of unsigned magnetic helicity created during a solar cycle is about $\approx 5 \times 10^{46}$ Mx$^2$ for solar cycle 22 \citep{Berger00} so only a factor 2 larger than the above estimate for MCs.  The amount of helicity created by the $\alpha$ effect is more difficult to estimate, but they argued that the amount is comparable or larger than the amount provided by differential rotation. Then the solar dynamo is able to create 4 or more times the amount of unsigned helicity found in MCs.  If most CMEs have a FR, and taking into account the flatness of the FR cross-section, the amount of unsigned helicity transported by CMEs is similar to the one produced by the solar dynamo.

The magnetic field, amplified by the global dynamo, emerges mostly in ARs.
Improved local correlation tracking methods have been developed to derive the photospheric velocities.  From these measurements magnetic helicity fluxes are derived \citep[\eg\ see the review of ][]{Demoulin09c}.  The largest input of helicity in the solar atmosphere is detected during the emergence of ARs. $\Hmc$ is about a factor 4 larger than the amount of unsigned helicity injection, $\approx 0.6 \times 10^{46}$ Mx$^2$, found by \citet{Georgoulis09} in emerging ARs over solar cycle 23.  However, $\Hmc$ is about the value found by \citet{Yang12b}, $\approx 3.3 \times 10^{46}$ Mx$^2$, and half  
the value found by \citet{Zhang13b}, $\approx 5 \times 10^{46}$ Mx$^2$ for AR emergence during solar cycle 23.  

We conclude that, within the present uncertainties of magnetic helicity estimations, \sect{Total-F-H}, the amount of magnetic helicity sent in MCs/CMEs is compatible both with the amounts of helicity built by the solar dynamo and measured in emerging ARs.

\section{Conclusion} 
      \label{sect_Conclusion} 
 
The \insitu\ measurements only provide local information of the physical parameters along the spacecraft trajectory.  
These data are typically fitted by a FR model to estimate the FR properties in a local 2D cut orthogonal to the FR axis. 
Since explorations of the same MC by several spacecraft are rare, we used a statistical approach to derive the generic properties of MCs.  
Indeed, different MCs are crossed at different location along their axis, providing statistical information along the axis. \citet{Janvier13} developed this new technique and they found the generic shape of MC axis only parametrised by the angular extension $\phimax$.  The location of the spacecraft crossing along the axis is related by the location angle $\lA$ which is available for each MC from the axis direction determined by fitting a FR model to the \insitu\ data.

In the present study, we further develop this statistical method by first analy\-sing how FR quantities vary along the FR axis and second by computing the axis length.  We find no dependence of the FR radius, field strength and twist along the FR axis in a broad range around its apex, within $|\lA|\leq 50 \degree$, for MCs observed at 1~AU.  The variation found for larger $|\lA|$ values are interpreted as a bias introduced when the spacecraft cross the FR legs and explore a significant portion of the FR along its curved axis while the fitted model has a straight axis.
We find a mean axis shape nearly symmetrical on both sides of the apex, with a very little asymmetry qualitatively consistent with the expected deformation given by the solar rotation.  Next, we derive the length of the generic axis:  
$1.3 \pm 0.6$~AU where the uncertainty is derived from the range of $\phimax$ observed for limb CMEs \citep{Wang11}.  If the FR is still attached to the Sun, the minimum length is $2.6 \pm 0.3$~AU.  The results allow the transformation of the local estimation of magnetic helicity, per unit length, to the total helicity of the FR.

The above results were applied to four lists of events: two with only MCs, one with MCs and small FRs and one with only small FRs.
While the small FRs largely dominate MCs in number (taking into account the probability to detect a FR), they have a much lower contribution (at least by a factor 10) than MCs for magnetic flux and helicity.  Indeed, MCs transport an important amount of magnetic flux and helicity when estimated over the full 1~AU sphere, as follows.  During one year, MCs carry a magnetic flux $F_{\rm MC} \approx 27 \times 10^{22}$ Mx$^2$ and an unsigned magnetic helicity $H_{\rm MC} \approx 2.3 \times 10^{45}$ Mx$^2$.  These are equivalent of the flux contained in about 25 large ARs and the equivalent of helicity injected in 200 emerging and large ARs (with a magnetic flux of $\approx 10^{22}$ Mx).  If all ICMEs possess a FR component, these numbers have to be multiplied by about a factor 3.   Finally, the amount of unsigned magnetic helicity carried away from the Sun by MCs during a solar cycle is comparable to the amount estimated for the solar dynamo and to the one measured in emerging ARs.

While we have improved the helicity estimation in MCs by analysing the FR parameter dependence along the axis and estimating the FR length, there are still a number of issues which require improvements.  A first one is the local FR model used to fit the magnetic data.  Does this model characterise well enough MCs?  Comparable helicity values were found with different models \citep{Gulisano05}, still it would be worth to do broader explorations both in terms of MCs and models, especially since doubts on the relevance of the Lundquist's model have recently come out \citep{Hu14}.   Second, FRs erode as they propagate in the solar wind \citep{Dasso06,Ruffenach15}. A deeper analysis of present data would allow to estimate both the helicity remaining in the FR at 1~AU and the one present before reconnection.  Third, many FRs do not have a circular cross section, so that an effort to fit \eg\ an elliptical model to the magnetic data would improve the helicity estimation.  Finally, the solar helicity \p{budget can be determined over the same time interval during which the MC helicity budget is studied.}  We conclude that there is a real potential to further improve our knowledge of MCs and in particular the solar magnetic helicity budget.       

\appendix   

\section{Mean Twist of a Flux Rope} 
    \label{sect_Mean-Twist}

We consider in this section any magnetic field with a cylindrical symmetry, so 
$\vec{B}(r)= \Ba(r) ~\ea + \Bz(r) ~\ez$, where $\Ba , \Bz$ are the azimuthal and axial components depending only on the radial coordinate $r$.
For this FR configuration, the magnetic helicity is linked to the amount of turns [$n(r)$] per unit length as shown below.  This is a concrete application of the more general expression of Equation~(12) of \citet{Berger-Field84} with poloidal/toroidal decomposition of a magnetic field.  However, the derivation rather follows the work of \citet{Berger03b} with the mutual helicity of ``open'' field (here $\Bz$) and ``closed'' field (here $\Ba$) which is gauge invariant.

The magnetic helicity, \eq{HL}, involves the vector potential component $\Aa$. 
Since $\curl \vec{A} = \vec{B}$,  $\Aa$ is linked to $\Bz$ as \citep[Equation (2) of][]{Dasso05b}      
  \begin{equation}  \label{eq_Aa}
  r~\Aa (r) = \int_{0}^{r} r' \Bz(r') \rmd r'= \frac{F(r)}{2 \pi}  \,,
  \end{equation}
where $F(r)$ is the axial magnetic flux within the circle of radius $r$. 
Writing the field line equations, the number of turns $n(r)$ is a function $\vec{B}$ components as
  \begin{equation}  \label{eq_n(r)}
  n(r) = \frac{\Ba(r)}{2 \pi ~r ~\Bz(r) }  \,.
  \end{equation}
Inserting \eq{Aa} in the left equality of \eq{HL} and replacing $\Ba(r)$ using \eq{n(r)} implies
  \begin{equation}  \label{eq_H-F-Ba}
  \del H = 2 \del L \int_{0}^{R} F(r) \, n(r) \,2 \pi \, r \, \Bz(r) \rmd r
           = 2 \del L \int_{0}^{R} F(r) \, n(r) \frac{\rmd F(r)}{\rmd r} \rmd r \,, 
  \end{equation}
where $L$ is the length and $R$ the radius of the FR.
Then, $\del H$ is rewritten as an integral on the axial flux $F$ as
  \begin{equation}  \label{eq_Hn}
  \del H = 2 \del L \int_{0}^{F} n(F') F' \rmd F'\,. 
  \end{equation}
When $n$ is independent of the radius $r$, so of $F(r)$, \eq{Hn} reduces to 
  \begin{equation}  \label{eq_HnUniform}
  \del H = n ~F^2 \, \del L \,. 
  \end{equation}
More generally, one defines $\nt$ with \eq{ntL} as a flux weighted mean of the number of turns per unit length.  Then, the magnetic helicity of a FR is always of the form of  \eq{HnUniform} with $n$ replaced by $\nt$, as written in \eq{Hnt}.

\begin{acknowledgements} \p{We thank the referee for her/his comments and enthusiasm on the manuscript.}
The data used in the present paper are provided by \citet{Lynch05}, \citet{Feng07,Feng08}, and \citet{Lepping10} also at 
http://wind.nasa.gov/mfi/mag\_cloud\_S1.html.
We thanks these authors for making those data publicly available.
This work was partially supported by the Argentinean grants 
PICT-2013-1462, UBACyT-20020120100220, and PIP-CONICET-11220130100439CO, and by a one month invitation by Paris Observatory of S.D.
S.D. is member of the Carrera del Investigador Cien\-t\'\i fi\-co, CONICET.
\end{acknowledgements}

   
\bibliographystyle{spr-mp-sola}

\bibliography{mc}  

\IfFileExists{\jobname.bbl}{} {\typeout{}
\typeout{****************************************************}
\typeout{****************************************************}
\typeout{** Please run "bibtex \jobname" to obtain} \typeout{**
the bibliography and then re-run LaTeX} \typeout{** twice to fix
the references !}
\typeout{****************************************************}
\typeout{****************************************************}
\typeout{}}



\end{document}